\documentclass{article}
\usepackage[utf8]{inputenc}
\usepackage{authblk}

\title{Quantum Heterodyne Sensing of Nuclear Spins via Double Resonance}
\author[1,2]{Jonas Meinel}
\author[1,2 ]{Minsik Kwon}
\author[1]{Durga Dasari}
\author[3]{Hitoshi Sumiya}
\author[4]{Shinobu Onoda}
\author[5]{Junichi Isoya}
\author[1]{ Vadim Vorobyov\footnote{v.vorobyov$@$pi3.uni-stuttgart.de}}
\author[1,2]{Jörg Wrachtrup}
\date{}

\affil[1]{3rd Physical Institute, Stuttgart, Germany}
\affil[2]{ Max Planck Institute for Solid
State Research, Stuttgart, Germany}
\affil[3]{Advanced Materials Laboratory, Sumitomo Electric Industries Ltd., Itami, Japan}
\affil[4]{Takasaki Advanced Radiation Research Institute, National Institutes for Quantum and Radiological Science and Technology, Takasaki, Japan}
\affil[5]{Faculty of Pure and Applied Sciences, University of Tsukuba, Tsukuba, Japan}

\usepackage{graphicx}
\usepackage{braket}
\usepackage{color}
\usepackage{dsfont}
\usepackage{bm}
\usepackage{amsmath,amssymb}
\usepackage{hyperref}
\usepackage{times}
\usepackage[normalem]{ulem}
\usepackage{multicol,caption}

\usepackage[square, compress,sort]{natbib}
\setcitestyle{numbers}
\usepackage{blindtext}
\usepackage{hyperref}
\usepackage[mathlines]{lineno}
\begin{document}
\setcitestyle{numbers}
\maketitle

\maketitle
\section*{Abstract}
Nanoscale nuclear magnetic resonance (NMR) signals can be measured through hyperfine interaction to paramagnetic electron sensor spins. 
A heterodyne approach is widely used to overcome the electron spin lifetime limit in spectral resolution. 
It uses a series of modified Hahn echo pulse sequences applied coherently with precession signal resulting in a subsampled NMR signal. 
Due to challenges with applying high electron Rabi frequencies its application is limited to low fields, thus the full potential of the method is not yet exploited at high magnetic fields, beneficial for NMR. 
Here we present heterodyne detection utilizing a series of phase coherent electron nuclear double resonance sensing blocks which extends nanoscale NMR protocols to arbitrary magnetic fields.
We demonstrate this principle on a single NV center, both with an intrinsic $^{14}$N and a weekly coupled  $^{13}$C nuclear spin in the bath surrounding single NV centres. 
We compare our protocol to existing heterodyne protocols and discuss its prospects. 
This work paves the way towards high field nanoscale heterodyne NMR protocols with NV centres which is crucial for reducing sample volumes and improving chemical resolution. 

\begin{multicols}{2}
\subsection*{Introduction} 
Nuclear magnetic resonance (NMR) is one of the most powerful analytical tools in medicine, chemistry, material science and physics \cite{Ernst1992}. 
NMR experiment measure small shifts of the Larmor frequency $\omega_L$ of a nuclear species allowing for resolving the chemical structure of molecules and solids, as well as their behaviour in various conditions and reactions. 
NMR signals are usually taken from millimetre sized objects, due to the limited sensitivity of the method. 
Various attempts have been made to measure the NMR signal of nanosized samples \cite{casey2020promise}.
One of the promissing avenues is the use of optically addressable paramagnetic centres, e.g. NV center in diamond as a sensor. 
High fidelity readout and long coherence times provide an excellent platform for sensing.
The optically detectable electron spin allows for single electron spin readout at room temperature.
Such single electron spin or a small ensemble of electron spins could be used as nano scale probes of local environments via observation of (modified) electron spin echo modulation signals (ESEEM), e.g. using dynamical decoupling sequences such as Carr-Purcell-Meiboom-Gill (CPMG), Knill dynamical decoupling in variations  \cite{suter}. 
However, in all of these techniques the intrinsic electron spin relaxation time $T_1$ is a fundamental limit to the achievable spectral resolution.

To overcome the intrinsic decay of the electron spin $T_1$ time, measurements can be performed sequentially, reinitialising the electron spin, and performing repetitive measurements. 
By using the linear regime of the sensor response, a phase sensitivity to the harmonic signal is achieved.  A sequence of measurements thus results in a subsampled harmonic signal \cite{schmitt2017submillihertz, boss2017quantum}. 
Each measurement is composed of electron spin initialisation, coherent interaction with the environment during the electron pulse sequence and optical electron spin readout. 
The pulse sequence is composed of MW $\pi$ pulses applied to the electron spin with periodicity of half the nuclear precession period. 
This technique is state of the art in high resolution methods denoted as quantum heterodyne (qdyne) or coherent averaged synchronised readout (CASR) \cite{glenn2018high}, which has demonstrated a sensor lifetime unlimited spectral resolution in detection of NMR and radio frequency signals.  
This technique has only been demonstrated at rather small fields $B < 0.1$ T where the nuclear Larmor precession frequency is still smaller than the available electron spin Rabi frequency (typically $\approx$10 MHz). 
While, conventional NMR experiments typically operate at a frequency range of $40-900 \, \mathrm{MHz}$.  
The higher thermal polarisation due to high field and the increase in spectral resolution beyond the diffusion limit \cite{schwartz2019blueprint}, makes it possible to detect chemical shifts and even J-coupling.
Current methods of NV sensing techniques applicable to high fields rely on in-situ correlation techniques such as coherent stimulated echo (CSTE) and 5-pulse ESEEM methods \cite{Laraoui2013,pfender2017nonvolatile,zaiser2016enhancing,aslam2017nanoscale}. 
In these methods the sensor interrogates with the nuclear environment twice, separated by a varied correlation time, which yields in a long time overhead and lacks the efficient sampling of the sinusoidal signals attributed to heterodyne techniques. 
Recently, the heterodyne sensing principle was applied to high frequency signals, e.g. in microwave band, by utilizing the interaction of an oscillating microwave field with the spin in the rotating frame, which in return was set by a stable external reference \cite{meinel2021heterodyne,chu2021precise,staudenmaier2021phase}. 

In this work, we utilize this effect for heterodyne detection of NMR signals at high magnetic fields. 
While for microwave detection we used its interaction with the spin for sensing its amplitude and phase, here we use a coherent radio frequency field to probe the nuclear spins  phase. 
This effectively down-converts the nuclear spins frequency to lower frequencies. 
For detection of this coherent response of the interaction we use a stimulated echo sequence with the NV center, which maps the z-projection of the nuclear spin  to the detectable fluorescence contrast of the NV center. 
%

\section*{Results}
\subsection*{Theory}
In this section we relate the evolution of the nuclear spin during our protocol with the observed electron spin polarization. 
The nuclear spin precesses freely with the Larmor frequency $\omega_L = \gamma_{\rm{N}} B$, proportional to the magnetic field $B$ by the gyromagnetic ratio $\gamma_N$. 
In the following, the nuclear spin and the sensor spin are described by the spin-$1/2$ operators $\hat{I}_{x,y,z}$ and $\hat{S}_{x,y,z}$ respectively. 
The nuclear precession is represented in the expectation value $\langle \hat{I}_x \rangle$ which needs to be encoded as a sensor phase via an effective Hamiltonian  $\hat{H}_{eff} \propto \hat{S}_z \otimes \hat{I}_x$. 
While previous work achieved this with dynamical decoupling \cite{Ma_2016}, which effectively removes the time dependence in the sensor-target effective Hamiltonian, we realise a coherent change of basis $\hat{I}_x \rightarrow \hat{I}_z$, $\hat{I}_y \rightarrow \hat{I}_y$, with a control RF driving field. 
After the rotation, the sensor can pick up the static component of the nuclear spin via the $\hat{H}_{eff} \propto \hat{S}_z \otimes \hat{I}_z$. 
Later, the nuclear spin is rotated back with transformation $\hat{I}_z \rightarrow \hat{I}_x$, $\hat{I}_y \rightarrow \hat{I}_y$. 
In the following we describe the heterodyne change of basis separate from the sensor target interaction. The Hamiltonians for each step are given by:
\begin{equation}
\begin{split}
	\hat{H}_{\rm{I}} &= \left(\omega_L - \omega_i\right) \hat{I}_z + \Omega_i(t) \hat{I}_x,\\
	\hat{H}_{\rm{II}} & = 2 \pi A_{zz}  \hat{S}_z \hat{I}_z,\\
\end{split}
\end{equation}
where $\Omega_i(t)$ is the rf driving field applied amplitude which is applied for the basis change, $\omega_i$ is the rf driving field frequency and $A_{zz}$ is the strength of the hyperfine interaction between sensor and target spin. 
$H_{\rm{I}}$ is responsible for the heterodyne response, analogous to \cite{meinel2021heterodyne,staudenmaier2021phase, chu2021precise} and summarized in the following, while $H_{\rm{II}}$ is used to measure the nuclear spin weakly. 

\begin{figure*}
\includegraphics[width=\textwidth ]{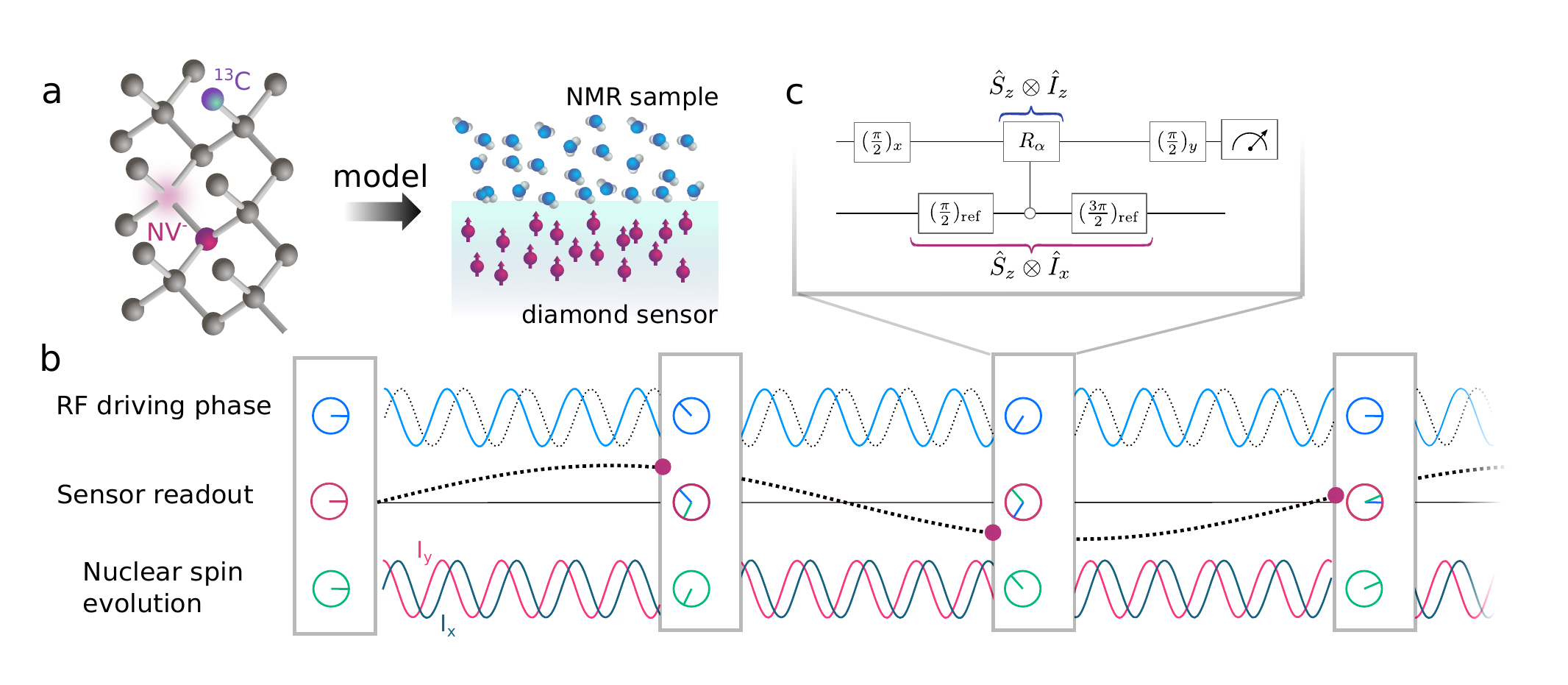}
\captionof{figure}{\label{fig:concept} \textbf{Concept of field independent heterodyne sensing by coherent double resonance.} \textbf{a} The NV center with its intrinsic $^{13}$C is studied as a model for a micron-scale surface NMR experiment with a diamond sensor. \textbf{b} The heterodyne interaction compares the phase of the RF driving source (blue pointer) to the phase of the nuclear spin evolution (green pointer) at each measurement $M_i$. In a series of measurments, the relative phase of the nuclear spin evolution can be reconstructed (red dots). \textbf{c} Conceptional circuit of a measurement $M_i$. The sensor is prepared in the superposition state with a $(\frac{\pi}{2})_x$ pulse and aquires a sensing phase of the target spin via the effective Hamiltonian of $\hat{S}_z \otimes \hat{I}_x$ before a second pulse $(\frac{\pi}{2})_y$ is applied and the population of the sensor is readout. The effective Hamiltonian is created by the coherent change in basis of the target spin relative to the RF driving phase ($x_{\rm{ref}}\rightarrow z$), followed by a static $\hat{S}_z \otimes \hat{I}_z$ interaction of the target and the sensor spin. Eventually, the target basis is again changed back to the original basis via interaction with the RF driving field ($z \rightarrow x_{\rm{ref}}$). The interaction with the sensor is static ($\hat{S}_z \otimes \hat{I}_z$) and therefore possible at arbitrary magnetic fields.}
\end{figure*}

From the initial nuclear spin state $\ket{\psi(0)} =  \ket{\uparrow}$ the superposition state $\ket{\psi(0)} =  \ket{\uparrow} - i \ket{\downarrow}$ is prepared with a $\pi/2$ pulse from the coherent RF control field $\Omega_i$. This state evolves under $H_{\rm{I}}$ relative to $\omega_i$ freely ($\Omega_i=0$) to:
\begin{equation}
	\ket{\psi(\tau)} = \ket{\uparrow} - i e^{i (\omega_L - \omega_i) \tau}\ket{\downarrow}.
\end{equation}
Further it is probed with a second $3\pi/2$ pulse from the same coherent source. 
This second pulse is phase sensitive and leads to the phase dependant expectation value:
\begin{equation}
	\langle I_z \rangle = \cos \left( \left( \omega_L - \omega_i \right) \tau \right).
\end{equation}
Instead of measuring $\langle I_z \rangle$ strongly, like in a Ramsey measurement \cite{pfender2017nonvolatile}, we measure this value weakly\cite{pfender2019high,cujia2019tracking}. 
Ignoring the measurement back action we can return the nuclear spin state with a $\pi/2$ pulse. 
The nuclear spin can again pick up a phase relative to the reference RF control field. For the $n$-th measurement we get:
\begin{equation}
	\langle I_z \rangle(n) = \cos\left( \left(\omega_L - \omega_i \right) n \tau \right).
\label{eq:heterodyne-response}
\end{equation}
This procedure can be repeated $N$ times until the nuclear spin state is dephased either naturally or via measurement back action. 
We compare the RF control field phase to the spin evolution phase which is the heterodyne response, illustrated in figure \ref{fig:concept}b.\\
In the following we describe the weak measurement with the electron spin via the system evolution under $H_{\rm{II}}$. 
The sequence is illustrated as a diagram in figure \ref{fig:concept}c, where the initialized sensor spin is prepared in $\left(\ket{0} + \ket{1}\right)/\sqrt{2}$ with a $\left(\pi/2\right)_x$ pulse. 
Then the sensor and the target spin evolve along the effective interaction term $\hat{H}_{eff} = 2 \pi A_{zz}S_z I_z$, see supplemental information \ref{supp:endor}, and is eventually brought to the population basis with another $\left(\pi/2\right)_y$ pulse. Finally the sensor spin population is read out. 
This encodes the heterodyne response, as in equation \ref{eq:heterodyne-response} in the sensor spin expectation value as:
\begin{equation}
	\langle S_z \rangle \left(n \right) = - \frac{1}{2} \sin\left(\alpha \right) \cos\left(\left(\omega_L - \omega_i\right) n \tau\right),
\end{equation}
where $\alpha$ is the weak measurement strength ($\alpha < 1$) given by $\alpha= \pi A_{zz}\tau_{zz}$ with $\tau_{zz}$ as the interaction time, i.e. the time between two $\pi/2$ pulses.  
The nuclear spin experienced a rotation of $2\pi = \pi/2 + 3 \pi/2$, neglecting weak measurement back action, and hence is already prepared for another iteration of the protocol. \\
In summary we straight forwardly applied the heterodyne sensing concept of microwave detection \cite{meinel2021heterodyne,staudenmaier2021phase, chu2021precise} inverted to the heterodyne sensing problem of a nuclear spin. 
The referencing is achieved by the control fields phase while the readout is realised through the static component in the hyperfine interaction between sensor and target spins and therefore independent of the Larmor frequency of the target spin.
This overcomes the field constraints of NMR with quantum sensors.

\subsection*{Experiment}
In the following, we experimentally test the sensing concepts with a single NV center measuring its intrinsic nuclear spins. 
The NV is at room temperature and can be optically initialized and readout. The microwave and radio frequency control is applied with a coplanar antenna making it a prime test bench for a proof of principle experiment.
The $^{14}$N, with its high readout fidelity \cite{neumann2010single}, acts as a test bed for the protocol's frequency response under experimental conditions. 
Additionally, the $^{12}$C enriched diamond lattice hosts residual single $^{13}$C atoms which couple to the single NV center. 
Because of the small coupling, the nuclear spin can be measured repeatately and can therefore simulate a weakly coupled distant spin bath. 
The interaction of the single $^{13}$C to the NV is given by the hyperfine interaction via a long range dipolar coupling of $H = \vec{S} \textbf{A} \vec{I} = 2 \pi A_{zx} S_z I_x + 2\pi A_{zz} S_z I_z$. 
We operate the setup at moderate magnetic fields of $B = 0.25 \, \mathrm{T}$, which allows us to access both components of the tensor.
Experimentally the $A_{zx}$ interaction is used to initialise the $^{13}$C nuclear spin via the polarisation exchange sequence (pulse-pol) \cite{schwartz2018robust}. 
While the $A_{zz}$ term is responsible for the signal acquired with our NV center via the proposed interaction. 
In a proof of principle experiment we demonstrate sequential weak measurements of a the $^{13}$C spin precession via coherent series of double resonance sequences. 
We further compare the spectral linewidth of the obtained NMR signal to the one obtained using conventional heterodyne experiments exploiting dynamical decoupling sensing blocks \cite{glenn2018high,schmitt2017submillihertz,boss2017quantum}.

\subsubsection*{Measurement of the heterodyne response with the intrinsic $^{14}$N nuclear spin}
\label{sec:freq-response}
\begin{figure*}
\centering
\includegraphics[width=\textwidth ]{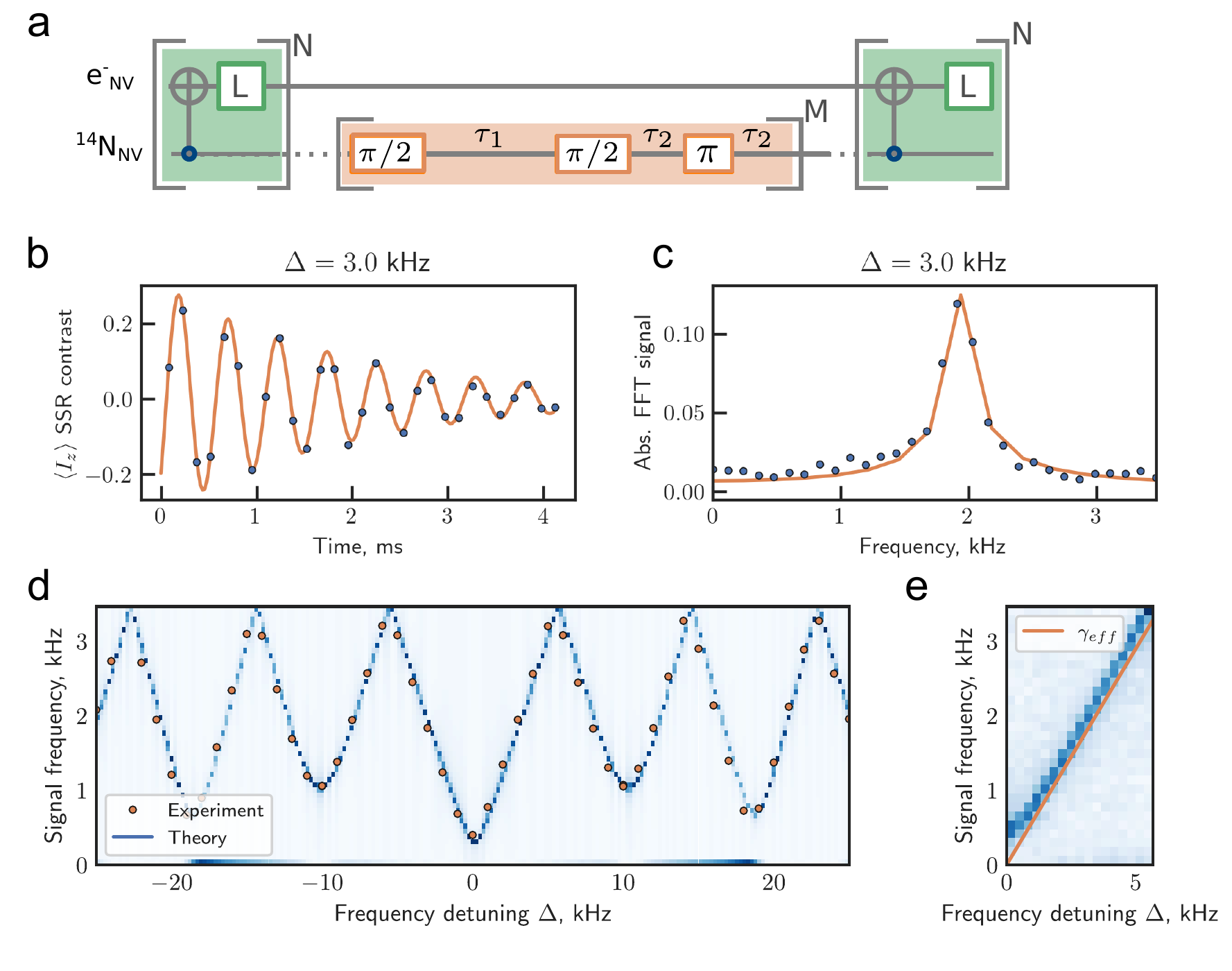}
\captionof{figure}{\label{fig:14N-frequency-response} \textbf{Heterodyne frequency response: projective measurments with the $^{14}$N intrinsic nuclear spin.} \textbf{a} The measurement protocol starts and ends with a single shot readout  through the electron spin with a laser pulse L. The initialized $^{14}$N spin is rotated by $\pi/2$ to start the precession for the time $\tau_1 = 83.333 \, \mu \mathrm{s}$ interrupted by a echo type rotation out of the precession plane with time $\tau_2=10 \, \mu \mathrm{s}$ responsible for the coherent change of basis. The time for a $2\pi$ rotation is $41 \, \mu \mathrm{s}$. \textbf{b} Measurement of the nuclear evolution. The rf-pulses where detuned by $\Delta = 3 \, \mathrm{kHz}$ relative to the nuclear resonance frequency $5.71435 \, \mathrm{MHz}$. \textbf{c} Fourier spectrum of the evolution of the same measurement. \textbf{d} Theoretical signal response spectra over $\Delta$. The spectra are overlayed with experimental obtained peak frequencies (blue dots spaced 1 kHz). A constant pulse error of 4 $\%$ was introduced to describes the spectrum at $\Delta = 0$. \textbf{e} Measured Fourier spectra over $\Delta$ for the first Brillouin-zone (from 0 to 6 kHz). The protocol shows a dominant linear response of the sensor frequency to the reference frequency. The linear relation is approximately $\gamma_{\mathrm{eff.}}=\tau_1/(\tau_1 + 2 \tau_2 + T_{\mathrm{rf}}) = 0.577$ plotted as an orange line in \textbf{e}. }
\end{figure*}

The NV centre's $^{14}$N nuclear spin $I=1$, which could be read out via single shot readout (SSR) \cite{neumann2010single}. SSR allows to initialise the $^{14}$N spin via measurement and strongly measure the spin state $\left(\ket{\uparrow}\; \mathrm{or} \;\ket{\downarrow}\right)$ at the end of the sequence thus estimating $\langle I_z \rangle$ by averaging over many experimental trials.  
We characterise the proposed protocol's simplified response, shown in equation \ref{eq:heterodyne-response}. We apply the rotation gates M times and estimate $\langle I_z \rangle$ at the end of the $M$-th rotation. The full experimental sequence is shown in figure \ref{fig:14N-frequency-response}a. The SSR consists of a CNOT-gate and an optical laser readout L, repeated $N=900$ times. 
Then we apply a $\pi/2$ pulse and bring the spin into x-y plane.
 We let the spin evolve in the x-y plane for a time $\tau_1$. 
 Afterwards, we rotate the spin out of the x-y plane with another $\pi/2$ pulse. 
 The following wait time $\tau_2$ simulates a sensor-target interaction. 
 We apply a $\pi$ pulse and another $\tau_2$ evolution time to echo out the phase evolution during the $\tau_2$ time, ensuring to suppress the phase pickup during the interaction.
 Finally, we rotate the spin back with a $\pi/2$ pulse, which already starts the second evolution for $\tau_1$. 
 The sequences repeats $M$ times. After the $M$-th repetition the nuclear spin projection is estimated with SSR. The frequency response is tested by changing the detuning $\Delta = (\omega_L - \omega_i)/(2\pi)$ of the rf-source used for the nuclear control pulses. 
 
Experimentally we implement this sequence with $\tau_1=83.333 \, \mu \mathrm{s}$, $\tau_2=10 \, \mu \mathrm{s}$ and the $T_\Omega = 41\, \mu \mathrm{s} $ for $M = 1,2,...,30$ repetitions. 
This leads to a total sequence time of $T_\mathrm{seq.}=144.333 \mu \mathrm{s}$. 
The result for $\Delta=3 \, \mathrm{kHz}$ is shown in figure \ref{fig:14N-frequency-response}b, where we show $\langle I_z\rangle$ over the experiment time $\left( M \cdot T_\mathrm{seq.}\right)$. 
The signal follows an exponential decaying sine with frequency $1.932\pm 0.003 \, \mathrm{kHz}$ and decay rate of $0.51 \pm 0.02 \, \mathrm{kHz}$. 
The decay rate, equal to a $T_2^*= 2.0 \pm 0.1 \, \mathrm{ms}$, is given by the electron spin lifetime $T_1=1.3 \pm 0.1 \, \mathrm{ms}$.  
In figure \ref{fig:14N-frequency-response}c we show the corresponding Fourier transformed signal amplitude.
For the input frequency of $\Delta =3 \, \mathrm{kHz}$ the signal frequency is reduced compared to the input. 
The relation between input and output signal is systematically studied in figure \ref{fig:14N-frequency-response}d, where we show the simulated frequency response of the sequence overlaid with experimental points spaced by $1 \, \mathrm{kHz}$.
We see good agreement between experiment and theory when accounting for a rf-pulse errors of $\epsilon=4 \%$ (see supplementary).
We sampled the first sector between $\Delta=0$ and $\Delta=6 \, \mathrm{kHz}$ with increased resolution of $0.333 \, \mathrm{kHz}$ shown in figure \ref{fig:14N-frequency-response}e. 
We see a largely linear response indicated by the reference line $\gamma_{\mathrm{eff.}} = \tau_1 / T_\mathrm{seq.}$. 
The effective response $\gamma_\mathrm{eff.}$ comes from the reduced precession time because of the rf-pulses and $\tau_2$ time overhead. 
The non linearities stem from the calibration errors of nuclear Rabi freqeuency $\Omega$, for small $\Delta$, and from to large rotation angle when the detuned Rabi frequency $\Lambda$ increases for large $\Delta$ as $ \Lambda =  \sqrt{\Omega^2+\Delta^2}$ . 
However, the frequency response is deterministic as the simulation demonstrates. 

\subsubsection*{Weak measurement of a single $^{13}$C nuclear spin}

\begin{figure*}
\includegraphics[width=\textwidth ]{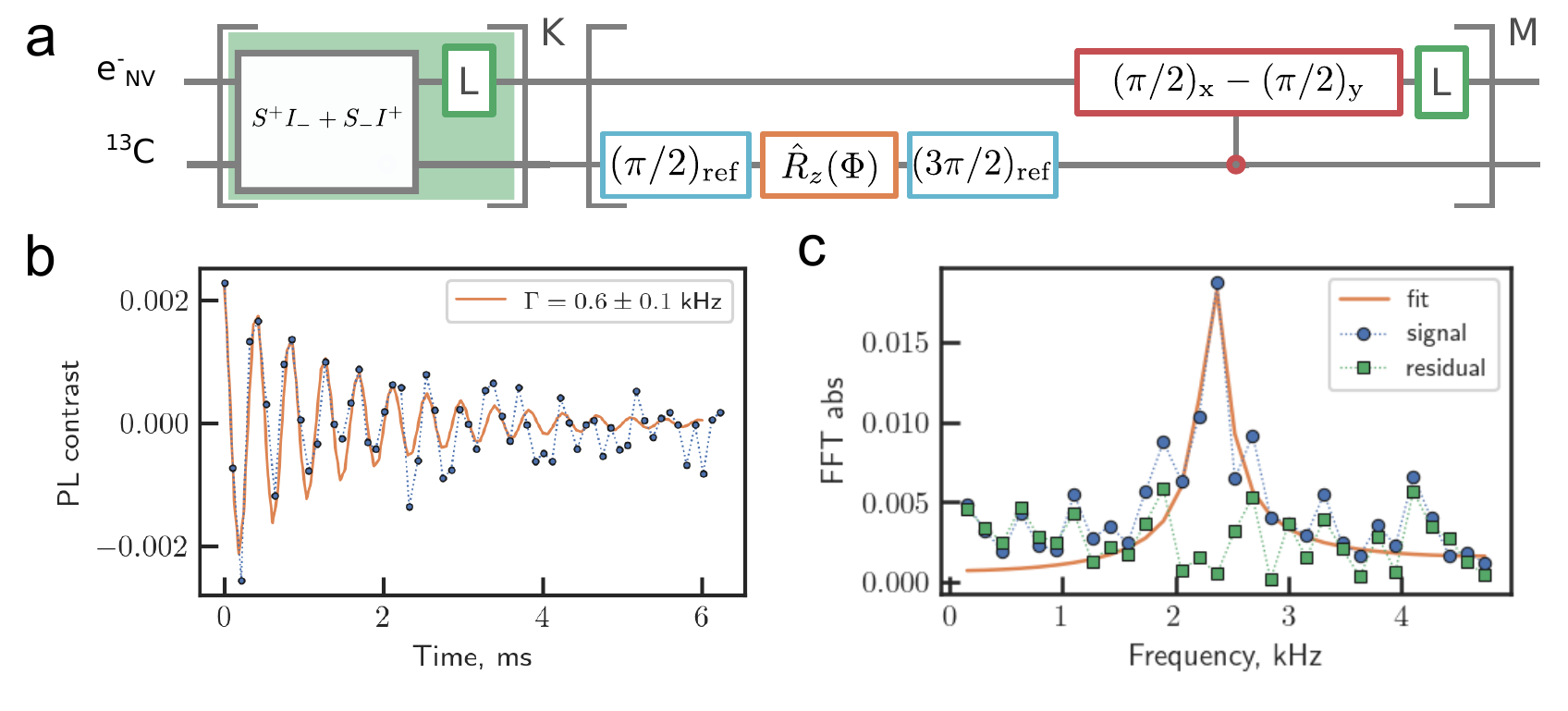}
\captionof{figure}{\label{fig:C13-proof-of-principle} \textbf{Proof of principle experiment with a single $^{13}$C spin and a single NV center.} \textbf{a} Sensing sequence of the proof of principle experiment. The $^{13}$C spin is hyperpolarized with the pulse-pol sequence, the nuclear spin precession is initiated with a $(\pi/2)_{\rm{ref}}$ pulse from the reference rf source. The nuclear spin evolves freely ($\hat{R}_z(\Phi)$) and gets rotated to the $z$-basis with a second $(3\pi/2)_{\rm{ref}}$ pulse from the same source. The z-component is measured with a Ramsey sequence on the sensor spin ($(\pi/2)_x - (\pi/2)_y$) and the sensor spin is readout with a laser pulse. Finally the nuclear spin is rotated back with a $(\pi/2)_{\rm{ref}}$ pulse and it follows again a free evolution. The sensing and free evolution are alternated $M$ times. \textbf{b} Experimental measurement of the nuclear evolution under the protocol above at resonance. For the proof of principle the $\hat{R}_z(\Phi)$ was implemented as a phase shift in the $3\pi/2$ pulse with $\Phi = \pi/2$, leading to a normalized signal periodicity of $4 \times  105.5506 \mu \mathrm{s}$. The decay rate of the signal is $\Gamma = 0.6 \pm 0.1 \, \mathrm{kHz}$. The spectrum of the signal with a peak at $2.368 \, \mathrm{kHz}$ is shown in \textbf{c}. Alongside the spectrum of the fit and the residuals of the fit from figure \textbf{b}.}
\end{figure*}

The residual abundance of $^{13}$C nuclear in the lattice allows to investigate individual weakly coupled nuclear spins in the $^{13}$C bath and probe them with our double resonance qdyne scheme in the regime of weak measurement. 
We  study a NV-$^{13}$C pair (NV$\#$5) with hyperfine coupling $A_{zz}=6 \, \mathrm{kHz}$. This pair was chosen as the coupling is sufficient to get a weak signal within $T_2^* = 50 \, \mu \mathrm{s}$ of the sensor while still being smaller than the nuclear Rabi frequency $\Omega_{^{13}\mathrm{C}} = 15 \, \mathrm{kHz}$. 
The later allows robust rf-control of nuclear spin with respect to NV center charge and spin state infidelities. The proof of principle experiment follows the sequence shown in figure \ref{fig:C13-proof-of-principle}a. 
The  $^{13}$C is initialised via the pulse-pol scheme \cite{schwartz2018robust} and is prepared in the superposition state with a $(\pi/2)_{\mathrm{ref}}$-pulse from the coherent RF source. 
After a free precession time, the spin is rotated to the measurement basis via a second pulse $(3\pi/2)_{\mathrm{ref}+\Phi}$, where the phase of the following pulses are cycled with a shift of $\Phi$ along the z-axis to move the demodulation frequency to 1/4 of the sampling frequency. 
The $^{13}$C $\langle I_z \rangle$ is mapped with a Ramsey sequence onto the NV sensor spin $\langle S_z \rangle$ which is optically readout. The $^{13}$C is rotated back with $(\pi/2)_{\mathrm{ref}+\Phi}$. 
The protocol is repeated $M$ times. In figure \ref{fig:C13-proof-of-principle}b the experimental result from the described sequence is shown. 
For this measurement the free precession time is $10$ $\mu$s, the interaction time (Ramsey) is $14.918$ $\mu$s, the nuclear Rabi period is $66$ $\mu$s which leads to a sampling time of $105.5506$ $\mu$s, when including the repolarisation time (1 $\mu$s) and wait times after rf-pulses (10 $\mu$s). 
The nuclear $^{13}$C signal is therefore at the expected frequency $2.368$ kHz, was measured to $2.336 \pm  0.018 \, \mathrm{kHz}$ and decays with a rate of $\Gamma = 0.6 \pm 0.1$ kHz. 
The Fourier transformed signal is shown in figure \ref{fig:C13-proof-of-principle}c, illustrating once more the frequency and related linewidth. 
In the following we analyze the obtained linewidth and compare it to the conventional qdyne measurements. 

\subsubsection*{Linewidth and Projected Spectral Resolution}
\begin{figure*}
\includegraphics[width=\textwidth ]{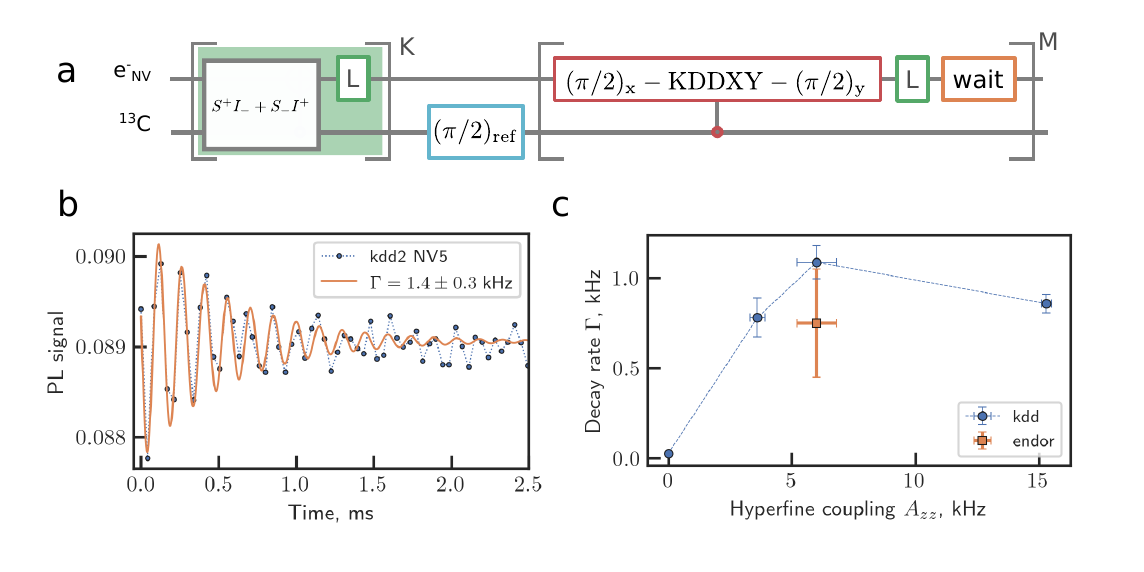}
\captionof{figure}{\label{fig:Linewidth-discussion} \textbf{Decay rate comparison to conventional DD-based heterodyne detection.} \textbf{a} Conventional DD-based heterodyne detection protocol \cite{cujia2019tracking, cujia2021parallel}. The $^{13}$C spin is hyperpolarized with the pulsepol sequence, the nuclear spin precession is initiated with a $(\pi/2)_{\rm{ref}}$ pulse from the reference rf source. The nuclear spin evolves freely, only interrupted by the measurement KDDXY sequence by  the sensor spin. The sensor is readout with a laser pulse L and a wait time is added to adjust the total sequence time. This block is repeated $M$ times. \textbf{b} Conventional qdyne signal from the sequence above with a decay rate of $\Gamma = 1.4 \pm 0.3$ kHz. In \textbf{c}, comparison of measurement back-action corrected lifetimes for different NV-$^{13}$C pairs with conventional qdyne (blue) to the ENDOR qdyne (orange). The lowest decay rate of $27 \pm 9 \, \mathrm{Hz}$ appears in the NV-$^{13}$C pair with $A_{zz} \leq  70 \, \mathrm{Hz}$.}
\end{figure*}

Magnetic resonance relies on high spectral resolution. 
Typically, qdyne experiments demonstrate the high spectral resolution with sub Millihertz resolution\cite{schmitt2017submillihertz}. 
Our qdyne experiment relies on a modified ENDOR experiment, therefore we must discuss the spectral resolution of our protocol in an electron nuclear interaction framework. 
In the following we quantify our spectral resolution in comparison with conventional dynamical decoupling based qdyne (conventional qdyne) and argue that the spectral resolution can be understood from that comparison.  

As a model system, we study the interaction of a single $^{13}$C with the NV electron spin. 
The interaction between electron and nuclear spin is summarised by a static $A_{zz}$ component and the oscillating components of $A_{zx}$. At high magnetic fields the effect of $A_{zx}$ is largely suppressed and hence we focus on the decoherence based on $A_{zz}$. 
In an experimental comparison between ENDOR qdyne to the conventional qdyne, we investigated  NV-$^{13}$C pairs with different $A_{zz}$ coupling. \newline
In figure \ref{fig:Linewidth-discussion}a the comparison sequence is shown analogous to references \cite{pfender2019high,cujia2019tracking}. 
We polarise the $^{13}$C spin with pulse-pol \cite{schwartz2018robust} polarisation transfer sequence and start the nuclear precession with a $(\pi/2)$ RF pulse. 
The repetitive sensing block consists of a Knill pulse sequence (KDDXY), an optical readout and a wait time. 
The wait time is crucial as in our proof of principle experiment we have an additional time due to RF pulses ($66 \, \mu \mathrm{s}$).
In figure \ref{fig:Linewidth-discussion}b, we compare the conventional qdyne to our ENDOR qdyne. 
The decay constant of the conventional qdyne was measured to be $\Gamma = 1.4 \pm 0.3$ kHz. 
The decay originates from the measurement back-action and intrinsic decay. 
Because the measurement back-action scales with the number of measurements it can be corrected when considering a series of measurements with increasing wait times, see supplementary information \ref{supp:decay-rates-exp}.
Further, with the conventional qdyne we can study the decoherence for $A_{zz}$ across a large range. In figure \ref{fig:Linewidth-discussion}c, we compared four NV-$^{13}$C pairs and find that the three NV-$^{13}$C pairs with $A_{zz}>3 \, \mathrm{kHz}$ show a decay rate of about 1 kHz, while only the NV-$^{13}$C pair with $A_{zz} \leq 0.07 \, \mathrm{kHz}$, also discussed in references \cite{meinel2021quantum,vorobyov2021robust}, shows a decay rate of $27 \pm 9 \, \mathrm{Hz}$. 

There are multiple causes for decoherence based on $A_{zz}$. 
First, the decoherence because of an interaction between the $^{13}$C with the NV center in the excited state during laser illumination \cite{chu2021precise} comes in. 
Next, the charge state switching between measurements is almost unavoidable, where the decoherence mechanism associated with NV$^0$ charge state electron spin is discussed e.g. in reference \cite{pfender2017nonvolatile}. 
Lastly, the initialisation infidelity of the NV$^-$ electron spin state\cite{song2020pulse} can shorten the lifetime \cite{maurer2012room} of the nuclear spin coherence. 
However, the exact combination of mechanisms causing the decoherence is still debatable and requires further research. 
The first two mechansims scale with $(A_{zz})^2$, as they originate from a random walk, and therefore can not explain the observed plateau. 
While only the last one, the decoherence based on spin initialisation infidelity, is in the right order of magnitude ($\approx 1 \, \mathrm{kHz}$) but we can not definitively pinpoint it, see supplementary information \ref{supp:infidelity}.

Independent from the exact description of the mechanisms for decoherence, the comparison between conventional qdyne and ENDOR qdyne yields that the spectral resolution limits are similar. 
However, with dynamical decoupling based qdyne, which gives the acces to $A_{zx}$ the extreme case of small $A_{zz} \rightarrow 0$ can be tested, resulting in a sensor unlimited spectral resolution below $30 \, \mathrm{Hz}$. 
This outlines the scenario where ENDOR qdyne could be advantageous when applied to small ensembles of nuclear spins. As already laid out in conceptual figure \ref{fig:concept}a, a distant ensemble of very weakly coupled nuclear spin ensemble is desirable for high spectral resolution. 
For example, an $A_{zz}$ coupling below 70 Hz is reached when the $^{13}$C $\left(^1 \mathrm{H}\right)$ is $\approx 4.5$ nm ($\approx 7$ nm) away from the NV center along the z axis, see supplementary information figure \ref{fig:distance}. 
The loss of coupling strength and therefore signal size has to be compensated by sensing an ensemble of nuclear spins with ensemble of NV centers, as already applied with conventional qdyne in reference \cite{glenn2018high,liu2022surface}.

\subsubsection*{Balancing sensitivity and magnetic susceptibility}

Beyond the proof of principle experiment we discuss the trade-off between the sensitivity of the sensor and the chemical information represented by the effective gyromagnetic ratio. 
Our protocol relies on two distinct parts: 
\begin{enumerate}
\item sensing the magnetic field from the nuclear spins,
\item free evolution of the nuclear spins.
\end{enumerate} 
It is clear that there must be a trade-off between them. To optimize the frequency estimation uncertainty, we calculated the Cramer-Rao lower bound of a decaying classical sinusoidal signal\cite{gemmel2010ultra} (see supplemental note \ref{supp:sensitivity}) given by:
\begin{equation}
	 \sigma(\nu) = \frac{2\sqrt{2}}{2\pi (A/\rho_\alpha) (T_2^*)^{3/2}},
\end{equation}
where $(A/\rho_\alpha)$ is the signal amplitude $A$ over the sensor noise spectral density $\rho_\alpha$ (units of $[\mathrm{T/\sqrt{Hz}}]$) and $T_2^*$ is the decoherence time of the nuclear spin target. 
The decoherence time $T_2^*$ from the nuclear spin bounds the advantageous power of $(-3/2)$ scaling \cite{schmitt2017submillihertz}.
In NMR spectroscopy, the chemical information is encoded in the frequency. 
Because of the control pulses and the static sensor target interaction, our protocol has a reduced frequency response of $\gamma_{\mathrm{eff.}}$, shown in figure \ref{fig:14N-frequency-response}. Therefore the chemical uncertainty scales inversely with $\gamma_{\mathrm{eff.}}$. The sensor noise spectral density $\rho_\alpha$ improves with the square root of the number of measurements in a given time. We find that the optimal ratio between free evolution and measurement time is (2:1) equivalent to $\gamma_{\mathrm{eff.}}=2/3$, see supplemental note \ref{supp:optimal-ratio}.\newline
After determining the optimal ratio, we can estimate the sensitivity of ENDOR qdyne. Our protocol relies on sensing of static fields \cite{zhang2021diamond,wolf2015subpicotesla,barry2016optical}, and in an improved version on an ENDOR type double resonance sequences\cite{mamin2013nanoscale}. The later can be applied when the nuclear spin $\pi$ pulses are shorter than the sensor spin echo coherence time $T_2$. Because of the time overhead, the typical sensitivities must be scaled accordingly by $\sqrt{T_{\mathrm{total}}/T_{\mathrm{sens.}}}$, where $T_{\mathrm{total}}$ is the sum of the nuclear spin manipulation time ($\approx \Omega_{rf}^{-1}$), the sensor spin preparation, sensing and readout time as well as the optimized free precession time. We estimated a sensitivity of $240\, \mathrm{nT/\sqrt{Hz}}$ for single shallow NV center and $60 \, \mathrm{pT/\sqrt{Hz}}$ for ensemble of NV centers, based on an extrapolation using common, as well as best values, see supplemental note \ref{supp:sensitivity}.

\section*{Summary, Discussion and Outlook}
In Summary we presented a protocol that shows a heterodyne response via double resonance. 
The principle is working beyond conventional dynamical decoupling and therefore removes the limitation to low field NMR. 
We theoretically and experimentally characterized the response of the double resonance qdyne sequence. 
Further we showed a weak measurement of a single $^{13}$C nuclear spin with our protocol. 
The extracted linewidth we compared to the conventional qdyne methods and find that the linewidth is intrinsically limited. 
Finally we discussed the effective sensitivity of heterodyne sensing and the service costs involved. \newline
As it becomes clear from this work, the double resonance qdyne is particular interesting for the nano to micron scale NMR sensing scenario, outlined in figure \ref{fig:concept}a. 
In this scenario, an ensemble of nuclear spins is coupled to the NV sensor spins\cite{liu2022surface,glenn2018high}. 
This means that the linewidth is not affected by the hyperfine dynamics like in our demonstrator experiment and the same spectral resolution as conventional qdyne can be achieved. 
In future, ENDOR qdyne could be combined with homo- and heteronuclear decoupling sequences\cite{waugh1968approach} improving the spectral resolution for solids and dipolar interacting samples.
The removed constraints on the microwave power allow to apply this method to high magnetic fields, e.g. optical detected magnetic resonance with NVs at 8 T \cite{fortman2021electron}. 
A high magnetic field leads to an increased thermal polarization equal to larger signals or equivalently reduced sensing volumes\cite{schwartz2019blueprint}. 
Additionally, the sensitivity to chemical shifts increases, allowing to study proteins and other shorter lived molecules \cite{lovchinsky2016nuclear}.  \newline
In the preparation of the manuscript, we become aware of a similar concept using the driven nuclear spin to detect nuclear magnetic resonance \cite{casanova2022}.

\subsection*{Acknowledgements}
We acknowledge financial support by European Union’s Horizon 2020 research and innovation program ASTERIQS under grant No. 820394, European Research Council advanced grant No. 742610, SMel, Federal Ministry of Education and Research (BMBF) project MiLiQuant and Quamapolis, the DFG (FOR 2724), the Max Planck Society, and the Volkswagentiftung.



\bibliographystyle{unsrtnat}
\bibliography{references}

\newpage

\end{multicols}

\newpage

\begin{center}
\title{\centering \Large Supplementary Information for "Quantum Heterodyne Sensing of Nuclear Spins via Double Resonance"}
\maketitle
\end{center}

\appendix
\section{Experimental Setup}

\begin{figure}[ht]
\centering
\includegraphics[width = 0.7 \textwidth]{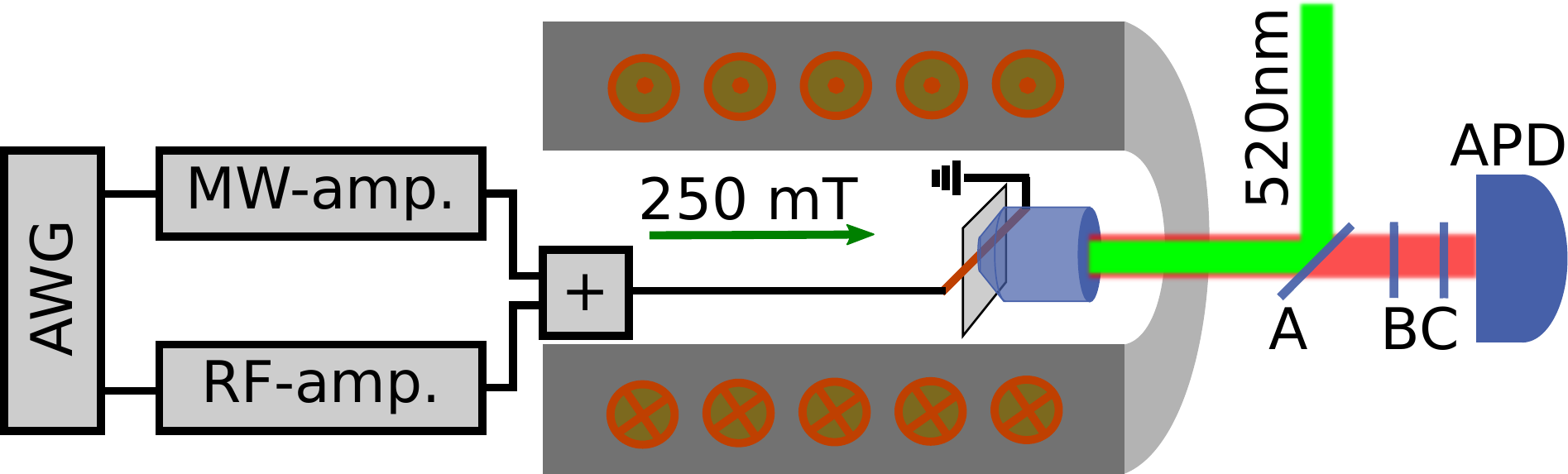}
\caption{ \textbf{Experimental setup}. The setup is a confocal room temperature single NV experiment. A 520 nm initialization and readout laser addresses the NV, the red fluoresence is passing the wedged mirror \textbf{A}, the pinhole \textbf{B} and the long pass filter 650 nm \textbf{C} before counted with the avalanche photo diode. The diamond is inside a room-temperature boar of a super-conducting magnet operated at around 250 mT. The piezo positioners and the microwave cables come in from the opposite side and connect to the coplanar wave guide where the diamond is glued. The microwave and radiofrequency control is generated from an arbitrary waveform generator (AWG) with two output channels. The mw-chanel is amplified in a traveling wave tube amplifier, while the rf signal is amplified in a solid state amplifier. They are combined before the sample and use the same coplanar waveguide to manipulate the nuclear and electron spin states.
\label{fig:experimental-setup}}
\end{figure}

The experiments were conducted with a confocal room temperature single NV setup depicted in figure \ref{fig:experimental-setup}. 
The NV is excited in the phonon sideband with a green 520 nm laser. 
This initializes and readouts the NV electron spin state. 
The red fluoresence of the NV is detected with an avalanche photo diode (APD) before passing through a wedged mirror, a pinhole $50 \,\mathrm{\mu m}$ and a long pass filter 650 nm. 
The diamond sample is a $2 \, \mathrm{mm} \times 2 \, \mathrm{mm} \times 80 \, \mu m$, (111)-oriented polished slice from a $^{12}\mathrm{C}$ -  enriched (99.995 \%) diamond crystal. 
The crystal was grown by the temperature gradient method under high-pressure high-temperature conditions at 5.5 GPa and 1350 $^\circ \mathrm{C}$, using high-purity Fe-Co-Ti solvent and high-purity $^{12}$C-enriched solid carbon. 
The single NV centers were created from intrinsic nitrogen by irradiaton with 2 MeV electrons at room temperature with a total fluence of $1.3 \cdot 10^{11} \, \mathrm{cm}^{-1}$ and annealed at $1000 \, ^\circ \mathrm{C}$ (for 2 h in vacuum). 
The typical lifetimes for the NV centers in this slice are $T_2^* = 50 \, \mu s$ and $T_2 \approx 300 \, \mathrm{\mu s}$  \cite{pfender2019high, meinel2021heterodyne}. 
The diluted $^{13}$C bath in the latice leads to a electron $T_2^*$ time of $50 \, \mu \mathrm{s}$ and usually only one $^{13}$C is significantly coupled to a single  NV center. 
The nuclear and electron spins are manipulated with an arbitrary waveform generator, able to sample 12 Gsamples/s, with two channels. 
The mw-channel is amplified with a Hughes-Traveling Wave Tube 8010H amplifier (TWT, max. 7 MHzRabi frequency), while the rf-channel is amplified with Amplifier Research 150A250. 
They are combined with a combiner before the both connect to the  coplanar waveguide where the diamond is glued to. We use a 3 axis piezo-positioner stage (range $100 \times 100 \times 25 \, (\mu \mathrm{m})^3$) from npoint. This allows us to keep a single NV in focus. \newline
For the proof of principle we searched for a single NV center with a moderate $^{13}$C with $A_{zz}$ coupling with a stimulated echo sequence \cite{pfender2017nonvolatile}. 
Moderate means that the coupling is smaller than the rf Rabi frequency for $^{13}$C of 15 kHz, while still being strong enough coupled to address it within the $T_2^*$ time. 
We found NV$\#5$ which has a coupling of $A_{zz}=6 \, \mathrm{kHz}$.
For the comparison of the natural linewidth we studied other NV centers within the $100 \times 50 \times 5 \, (\mu \mathrm{m})^3$ volume.

\section{Theoretical Description of the Heterodyne Double Resonance}
\label{supp:endor}
\subsection{Heterodyne response by coherent change of basis}
The heterodyne double resonance relies on the coherent change of basis followed by a weak measurement. We consider a polarized nuclear spin described by the spin operators $I_{x,y,z} = \frac{1}{2}\sigma_{x,y,z}$ with $\sigma_{x,y,z}$ the Pauli matrices and the nuclear eigenstates are $\ket{\uparrow}$ and $\ket{\downarrow}$. The initial density matrix $\rho_0=\ket{\uparrow}\bra{\uparrow}$ evolves under the Hamiltonian 
\begin{equation}
H = (\omega_L - \omega_i) I_z + \Omega_i(t)I_x,
\end{equation}
with $\Omega_i(t)$ reference pulse field amplitude and $\omega_i$ the reference field frequency. In the following we use $\delta \omega = (\omega_L - \omega_i)$. The reference field is applied until the nuclear spin is rotated by $\pi/2$ leading to:
\begin{equation}
\rho_1 = \frac{1}{2} (\mathds{1} - \sigma_y),
\end{equation}
the state evolves under $H = \delta \omega I_z$ to:
\begin{equation}
\rho_2 = \frac{1}{2}\begin{pmatrix}
1 & i\; e^{-\delta \omega \tau} \\
- i\; e^{\delta \omega \tau} & 1 
\end{pmatrix}
\end{equation}
afterwards we apply a $3 \pi/2$ pulse. We get:
\begin{equation}
\rho_3 = \frac{1}{2}\begin{pmatrix}
1 + \cos(\delta \omega \tau) & \sin(\delta \omega \tau) \\
\sin(\delta \omega \tau) & 1 - \cos(\delta \omega \tau) 
\end{pmatrix}
\end{equation}
and the expectation value:
\begin{equation}
\langle I_z \rangle = \frac{1}{2} \cos(\delta \omega \tau).
\end{equation}
This expectation value will be measured weakly, however we neglect the back-action for now. We rotate the state back with a $\pi/2$ pulse recreating state $\rho_2$, which will evolve under Hamiltonian $H$ for time $\tau$ and the sequence repeats. For a series of measurements we get the expectation values:
\begin{equation}
\langle I_z(n) \rangle = \frac{1}{2} \cos(\delta \omega \tau n),
\end{equation}
equal to the demodulation signal. \\

\subsection{Numerical Simulations with Pulse-Errors}

The deriviation of the above assumes the perfect linear response and illustrates the idea. However, it neglects pulse errors from the description. Here we consider only two simply errors, which was sufficient to describe the data shown in figure \ref{fig:14N-frequency-response}. The first one is a the off-resonance $\Delta$ of the rf-control pulses, the second a calibration error of the rabi frequency of $\epsilon$. The pulse is described by the evolution of this Hamiltonian:
\begin{equation}
	H = \Delta \, I_z + \Omega \, (1 + \epsilon) \, I_x.
\end{equation}
When we compute the unitary evolution operator of this Hamilitonian we define the $\pi/2 =\Omega \, \tau_{\pi/2}$ and $\pi = \Omega \, \tau_{\pi}$ pulse based on the orignal Rabi frequency $\Omega$. The figure \ref{fig:14N-frequency-response} shows the response for $\epsilon = 0.04$, $\Omega= 2\pi \cdot 25 \, \mathrm{kHz}$ and $\Delta$ is determined by the off-resonance shown on the x-axis of the figure. 

\subsection{Sensor target interaction}
Next we describe the interaction to the sensor spin $S_{x,y,z} = \frac{1}{2} \sigma_{x,y,z}$ (reduced to two level system of the NV). We start from the initial total density matrix with an initialized sensor spin brought to superposition to $\rho_s =  \frac{1}{2} (\mathds{1} - i \sigma_y)$ and the nuclear spin density matrix $\rho_{\mathrm{k}}$ summarized to $\rho_{\mathrm{total}} = \rho_s \otimes \rho_{\mathrm{k}}$ and explicitly given by: 
\begin{equation}
	\rho_{\mathrm{total},0} = \frac{1}{4}\begin{pmatrix}
1 & i\\
-i & 1 
\end{pmatrix}
\otimes
\begin{pmatrix}
1 + \cos( \beta) & \sin(\beta) \\
\sin(\beta) & 1 - \cos(\beta) 
\end{pmatrix},
\end{equation}
where we summarized  $\beta =\delta \omega \tau $.
This state evolves under the Hamiltonian $H_{II} = 2 \pi A_{zz} S_z I_z$ to:
\begin{equation}
	\rho_{\mathrm{total},1} = \frac{1}{4} \begin{pmatrix}
1 + \mathrm{c}(\beta) & \mathrm{s}(\beta) e^{-i\alpha/2}  & i(1 + \mathrm{c}(\beta))e^{-i\alpha/2} & i \mathrm{s}(\beta) \\
\mathrm{s}(\beta)e^{i\alpha/2} &  1 - \mathrm{c}(\beta) & i\mathrm{s}(\beta) &  i(1 - \mathrm{c}(\beta))e^{i\alpha/2} \\
-i(1 + \mathrm{c}(\beta))e^{i\alpha/2} & -i \mathrm{s}(\beta) & 1 + \mathrm{c}(\beta) &  \mathrm{s}(\beta)e^{i\alpha/2} \\
-i \mathrm{s}(\beta) &  -i(1 - \mathrm{c}(\beta)) e^{-i\alpha/2}  & \mathrm{s}(\beta)e^{-i\alpha/2} &   1 - \mathrm{c}(\beta) \\

\end{pmatrix}
\end{equation}
where $\alpha = \pi A_{zz} \tau_{\mathrm{int}}$, $\cos() = \mathrm{c}()$ and $\sin() = \mathrm{s}()$ were used to shorten the expression. Finally we rotate the sensor spin with a $\pi/2$ pulse along the y-axis to:
\begin{equation}
\frac{1}{4}\begin{pmatrix}
2(1 + \mathrm{c}(\beta))(1-\mathrm{s}(\alpha)) & 2 \mathrm{s}(\beta)\mathrm{c}(\alpha)  & 2 i(1 + \mathrm{c}(\beta))\mathrm{c}(\alpha) & 2 i \mathrm{s}(\beta) (1 - \mathrm{s}(\alpha)) \\
2 \mathrm{s}(\beta)\mathrm{c}(\alpha) &  2(1 - \mathrm{c}(\beta))(1+\mathrm{s}(\alpha)) & 2i \mathrm{s}(\beta) (1+ \mathrm{s}(\alpha)) &  2i(1 - \mathrm{c}(\beta))\mathrm{c}(\alpha) \\
-2 i(1 + \mathrm{c}(\beta))\mathrm{c}(\alpha) & -2i \mathrm{s}(\beta) (\mathrm{s}(\alpha)+1)& 2(1 + \mathrm{c}(\beta))(1+\mathrm{s}(\alpha)) &  2\mathrm{s}(\beta)\mathrm{c}(\alpha) \\
-2 i \mathrm{s}(\beta) (1 - \mathrm{s}(\alpha)) &  -2i(1 - \mathrm{c}(\beta))\mathrm{c}(\alpha)  & 2\mathrm{s}(\beta)\mathrm{c}(\alpha) &   2(1 - \mathrm{c}(\beta))(1-\mathrm{s}(\alpha)) \\

\end{pmatrix},
\end{equation}
when we take the expectation value $S_z$ of sensor spin from the total density matrix, we get:
\begin{equation}
	\langle S_z \rangle = -\frac{1}{2}\cos(\beta) \sin(\alpha) = -\frac{1}{2}\cos(\delta \omega \tau) \sin(\alpha).
\end{equation}
\subsection{Measurement back-action}
Finally we discuss the measurement back action analogous to reference \cite{pfender2018mitigating}. The measurement back action lead to two possible final density matrices of the nuclear spin:
\begin{equation}
\begin{split}
 \rho_\uparrow  &= \frac{1}{2} \begin{pmatrix}
(1 + \cos( \beta))(1-\sin(\alpha)) & \sin(\beta) \cos(\alpha) \\
\sin(\beta) \cos(\alpha) & (1 - \cos(\beta) )(1+\sin(\alpha))
\end{pmatrix},\\
\rho_\downarrow &=\frac{1}{2} \begin{pmatrix}
(1 + \cos( \beta))(1+\sin(\alpha)) & \sin(\beta) \cos(\alpha) \\
\sin(\beta) \cos(\alpha) & (1 - \cos(\beta) )(1-\sin(\alpha))
\end{pmatrix} 
\end{split},
\end{equation}
when we average over the measurement results, we find :
\begin{equation}
\frac{1}{2}(\rho_\downarrow + \rho_\uparrow) =\frac{1}{2} \begin{pmatrix}
1 + \cos( \beta) & \sin(\beta) \cos(\alpha) \\
\sin(\beta) \cos(\alpha) & 1 - \cos(\beta) 
\end{pmatrix} = \frac{1}{2} \mathds{1} + \cos(\beta) I_z + \sin(\beta) \cos(\alpha) I_x.
\end{equation}
Only the $I_x$ component is reduced independent of the measurement outcome. When we look at a series of measurements we can express it as a series of rotation of an angle $\beta$ and a measurement back-action of $\cos(\alpha)$:
\begin{equation}
	\begin{pmatrix}
\langle I_x (n+m)\rangle \\
\langle I_z (n+m) \rangle \\
\end{pmatrix} 
= \begin{pmatrix}
\cos( \alpha) \cos(\beta) & -\sin(\beta) \\
\cos( \alpha) \sin(\beta) &  \cos(\beta)
\end{pmatrix} ^m
\begin{pmatrix}
\langle I_x (n)\rangle \\
\langle I_z (n)\rangle \\
\end{pmatrix},
\end{equation}
where we calculate the $(n+m)$-th measurement from the $n$th measurement. We can simplify the above by diagonalizing the matrix with the eigenvalues:
\begin{equation}
	\lambda_{\pm} = \left( \cos(\beta) \pm i \sin(\beta) \sqrt{1 - \frac{\mu^2}{\sin^2(\beta)}} \right) \cos^2(\alpha / 2),
\end{equation} 
with $\mu = \tan^2(\alpha/2)$. For a weak measurement and regular demodulation frequencies ($\sin(\beta) \gg 0$) the square root is equal to 1 and we can simplify the expression to:
\begin{equation}
	\lambda_{\pm} = e^{-\Gamma_{\mathrm{eff}} \pm i \beta},
\end{equation}
with $\Gamma_{\mathrm{eff}} = \frac{\alpha^2}{4}$. Defining the starting conditions of $\langle I_x (0)\rangle = 0$  and $\langle I_z (0)\rangle = \frac{1}{2} $, with the simplifications of above,  we get:
\begin{equation}
	\begin{pmatrix}
\langle I_x (m)\rangle \\
\langle I_z (m)\rangle \\
\end{pmatrix} 
= \frac{1}{2}\begin{pmatrix}
\sin(\beta m)\; e^{-\Gamma_{\mathrm{eff}}m}\\
\cos(\beta m)\; e^{-\Gamma_{\mathrm{eff}}m}  \\
\end{pmatrix}. 
\end{equation}
The cases where these simplifications do not apply are analogous to conventional weak measurements and can be found e.g. in reference \cite{cujia2019tracking,pfender2019high}.\newpage

\section{Experimental Design and Data Analysis of Double Resonance qdyne with $^{13}$C}

\begin{figure}
\centering
\includegraphics[width = \textwidth]{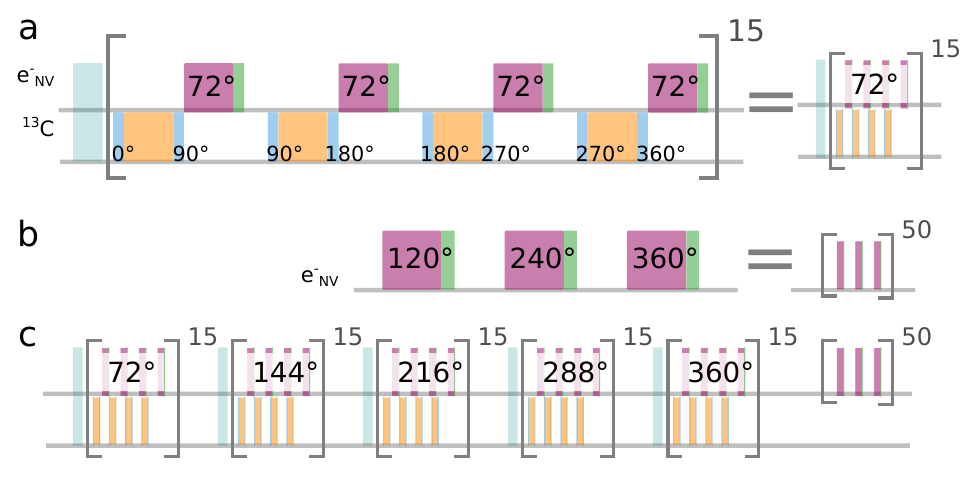}
\caption{ \textbf{Experimental design of the pulse sequence for double resonance qdyne}. \textbf{a} The protocol starts with a pulse-pol sequence (olive), then the $^{13}$C spin precession starts with a $\pi/2$-pulse of phase $0^\circ$ (blue), after $10 \, \mu \mathrm{s}$ (orange) free precession time a $3\pi/2$-pulse (blue) of phase $90^\circ$ follows rotating the spin out of plan. A Ramsey experiment on the electron sensor spin (purple), consisting of a $\pi/2$ ($0^\circ$), sensing time $\tau_2$, another $\pi/2$ ($72^\circ$) and finally an optical readout (green) follows. This measures the nuclear spin weakly. The nuclear spin is rotated back in the precession plane with a $\pi/2$ ($90^\circ$) pulse. The shifting of the $^{13}$C-$\pi/2$ pulses by $90^\circ$ moves the signal to $1/4$ of the sampling frequency. \textbf{b} Phase modulated Ramsey magnetometer sequences applied to measure the phase offset of the microwave pulse. \textbf{c} Final overall experimental sequence, the 5 implementations of the sequence shown in \textbf{a} create a dataset, which allows to correct the slow long term phase drifts with the data aquired from \textbf{b}.
\label{fig:sequence-ddq}}
\end{figure}

\begin{figure}
\centering
\includegraphics[width = \textwidth]{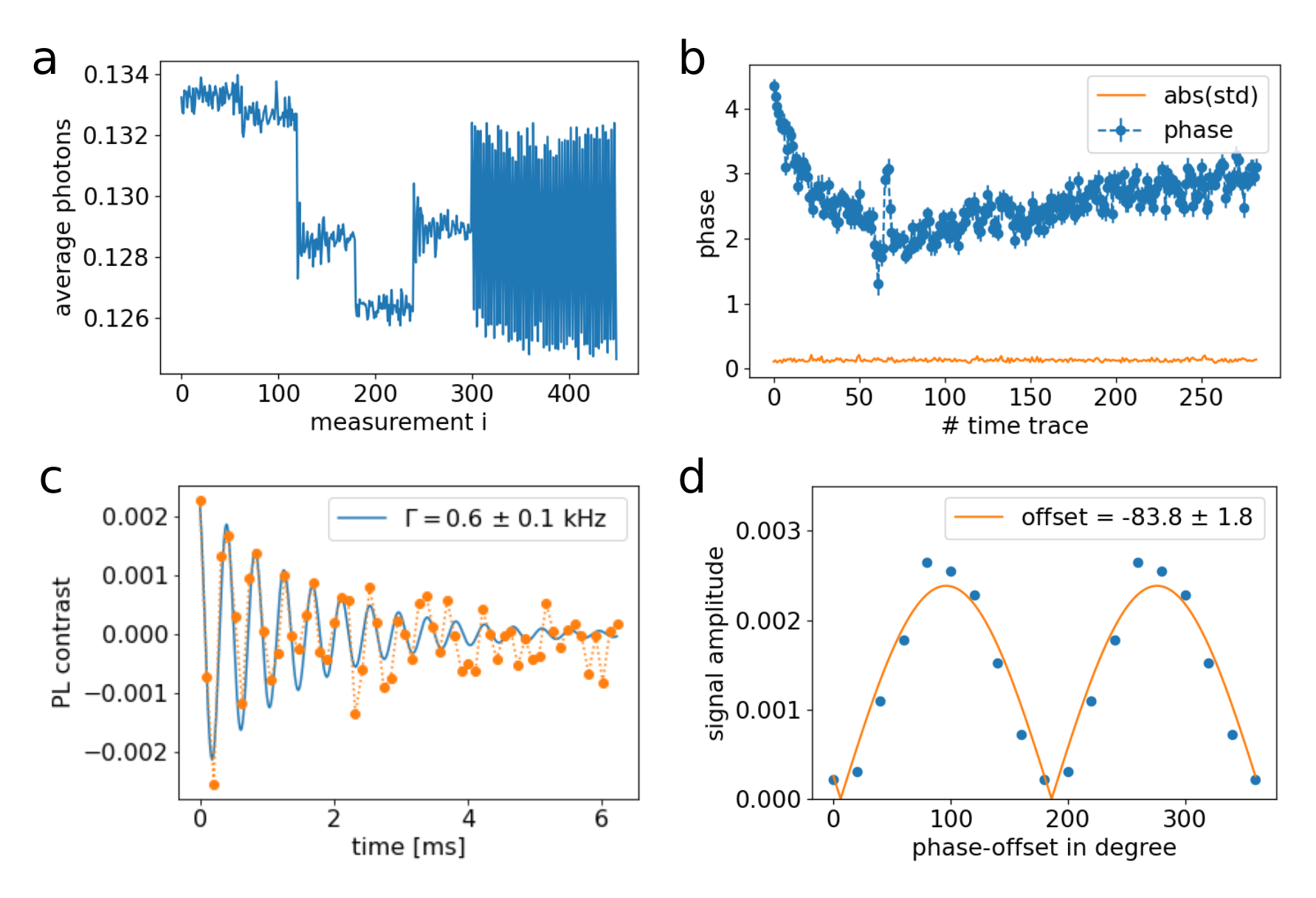}
\caption{ \textbf{Experimental raw data and data processing}. \textbf{a} Averaged photon counts over the full measurement sequence. The first 5 realizations of 60 measurements are the double resonance qdyne experiments, followed by 150 measurements for the sensor phase measurement. \textbf{b} Extracted phase value from the reference measurements. \textbf{c} The averaged signal grouped within the range of $60^\circ$ to $120 ^\circ$ around the central value of $90^\circ$. \textbf{d} Extracted signal amplitude when the central value was swept. 
\label{fig:raw-data-processing}}
\end{figure}

The double resonance qdyne relies on the sensor target interaction along the z-direction. It becomes mundane to make sure that the sensor picks up the nuclear signal while other signal, e.g. from temperature or magnetic field changes, are corrected. 
Here we discuss how we approached this challenge. 
We opt for two complementary passive methods which allow to later correct the dataset. 
First we implement the experiment with at least 4 to 5 readout angle realisations and second we measure the sensor phase after each block of experiments. \newline
In figure \ref{fig:sequence-ddq} a, we show the sequence of the double resonance. The sequence starts with the polarization of the nuclear target spin via pulse-pol (olive color). 
Then the $^{13}$C nuclear spin precession is started with a $(\pi/2)_{\mathrm{rf},0^\circ}$ pulse from our coherent radio frequency source (AWG). 
The nuclear spin precesses for $10 \, \mu \mathrm{s}$ and gets rotated with a $(\pi/2)_{\mathrm{rf},90^\circ}$. 
The nuclear polarisation relative to the coherent rf-source is measured with a Ramsey experiment (purple) on the NV-electron spin $(\pi/2)_{e,0^\circ} - (\pi/2)_{e,72^\circ}$. 
The NV population is measured with an optical readout pulse (green). 
The nuclear spin precession is continued with a $(3\pi/2)_{\mathrm{rf},90^\circ}$. The $90^\circ$ phase shifts in the reference pulses move the signal to 1/4 of the sampling frequency when cycled ($0^\circ \rightarrow 90^\circ \rightarrow 180^\circ \rightarrow 270^\circ \rightarrow  ...$) as shown in the lower panel of figure \ref{fig:sequence-ddq} a. 
We repeat this sequence for 15 times, resulting into 60 weak measurements. \newline
The NV-electron spin alongside the nuclear signal picks up the magnetic enviromental noise, leading to a change in readout angle. 
We measure this bias angle with an magnetometry sequence with a increased bandwidth, as in \cite{zhang2021quantum}. 
The electron spin magnetometry sequence is given by 
\begin{equation}
\left\{(\pi/2)_{e,0^\circ} - (\pi/2)_{e,0^\circ}, \, (\pi/2)_{e,0^\circ} - (\pi/2)_{e,120^\circ}, \, (\pi/2)_{e,0^\circ} - (\pi/2)_{e,240^\circ}\right\}.
\end{equation} 
The enviromental noise phase shift is extracted from the phase shift of the sinusoidal modulated photon counts. \newline
Finally the overall assembled sequence is shown in figure \ref{fig:sequence-ddq} c. 
It consists of 5 blocks of the sequence shown in figure \ref{fig:sequence-ddq} a with different readout angles, and a magnetometry sequence at the end of the sequence. \newline
The total sequence has 450 optical readouts. 
We typically measure for 1-2 million optical readouts, before we refocus the NV in the confocal spot. 
The detected binned photon counts can be averaged in bundles of 450 as shown in figure \ref{fig:raw-data-processing} a. 
We see the 5 angles of the double resonance qdyne experiment and the strong magnetometer signal at the end of the sequence. 
The magnetometer starting phase of each photon time trace is analyzed and the phase offset is extracted, shown in figure\ref{fig:raw-data-processing}b. 
The phase has stable periods however occasionally we have strong changes, e.g. at time trace number 50. 
However as we measure the multiple implementations of the experiment we can now decide which experiments to select and which to neglect based on the phase offset. 
This decision was made within a $\pm(90^\circ \pm 30^\circ)$ corrected phase acceptance window. 
The result after the grouping and averaging is shown in figure \ref{fig:raw-data-processing} c. 
The signal size depending on the central value of this acceptance window was analyzed in figure \ref{fig:raw-data-processing} d. 
We clearly see the signal having its maximal amplitude around $\pm 90^\circ$.

\newpage

\section{Extended Data: Linewidth Comparison}
\label{supp:decay-rates-exp}
\begin{figure}
\centering
\includegraphics[width = \textwidth]{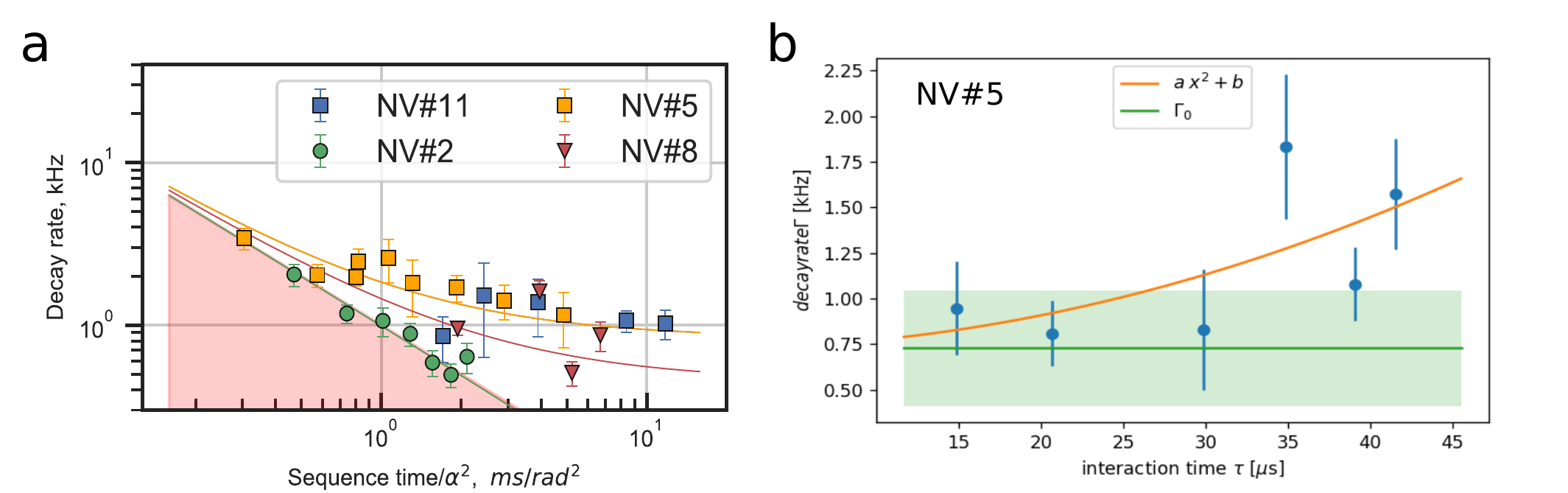}
\caption{ \textbf{Determining the natural decay time $\Gamma_0$}. \textbf{a} The decay rate for different NV-$^{13}$C pairs over the sequence time normalized by the measurement back-action $\alpha^2$. The red shaded area indicates the measurment back-action regime, the limit of the decay rate is $\Gamma_0$. \textbf{b} The decay rate $\Gamma_0$ in the double resonance qudnye experiment was measured for the NV-$^{13}$C pair NV$\#5$. Here the interaction time $\tau_2 \propto \alpha$ was increased to seperate the back-action from $\Gamma_0$.
\label{fig:decay-plots}}
\end{figure}

In this section we show the extended data that contributed to figure \ref{fig:Linewidth-discussion} in the main text. First we analyze the decay time from the conventional qdyne. We used the protocol shown in figure \ref{fig:Linewidth-discussion} a, where we polarize the $^{13}$C nuclear spin with pulse pol. Then we start the spin precession with a $(\pi/2)_{\mathrm{rf},0^\circ}$ pulse. During the free precession, the $^{13}$C spin is measured by weak entanglement with the sensor nuclear spin and optical readout. In previous work \cite{cujia2019tracking}, the decoherence mechanism was considered as largely happening during green laser illumination. However, here we want to study the decoherence when the green laser pulses are spread far apart like in our double resonance qdyne protocol. For this we introduce a wait time to increase the sequence duration to match the sequence length of the double resonance qdyne. For each NV-$^{13}$C pair we implement multiple experiments with increasing wait time to seperate the measurement back action decay from the natural decay. We parameterize the decay by:
\begin{equation}
	\Gamma_{\mathrm{total}}(\tau_{\mathrm{seq.}}) = \frac{\alpha^2}{4\, \tau_{\mathrm{seq.}}} + \Gamma_0,
\end{equation}
where $\tau_{\mathrm{seq.}}$ is the total sequence time, $\alpha^2/4$ parameterize the measurement back-action and $\Gamma_0$ is the natural linewidth under the protocol. In figure \ref{fig:decay-plots} a we show the decay time following the equation above. For comparability the of different NV-$^{13}$C pairs we plot the x-axis as $\tau_{\mathrm{seq.}} / \alpha^2$ destinctively seperating the measurement backaction dominant part (red) from $\Gamma_0$ (the asymptotic level). The fit results were used to create figure \ref{fig:Linewidth-discussion}c for the kdd-sequences. \newline
On the other hand we analysed the decay limit from the double resonance qdyne experiment. Because of the already very long sequence time, we here increased the measurement strength $\alpha \propto \tau_{zz}$ by increasing the interaction time $\tau_{zz}$ as shown in figure \ref{fig:C13-proof-of-principle} as sequence schematic. This allows to measure the measurement back-action free decay limit shown by the fit in figure \ref{fig:decay-plots} b. The green horizontal line indicates the limit value extracted from the fit and the green shaded area is the uncertainty of $\Gamma_0$. The value of $(.75 \pm 0.3) \, \mathrm{kHz}$ is entered in figure \ref{fig:Linewidth-discussion} c for the double resonance qdyne. 

\newpage

\section{Decoherence by initialization fidelity}
\label{supp:infidelity}
\begin{figure*}
\includegraphics[width=\textwidth]{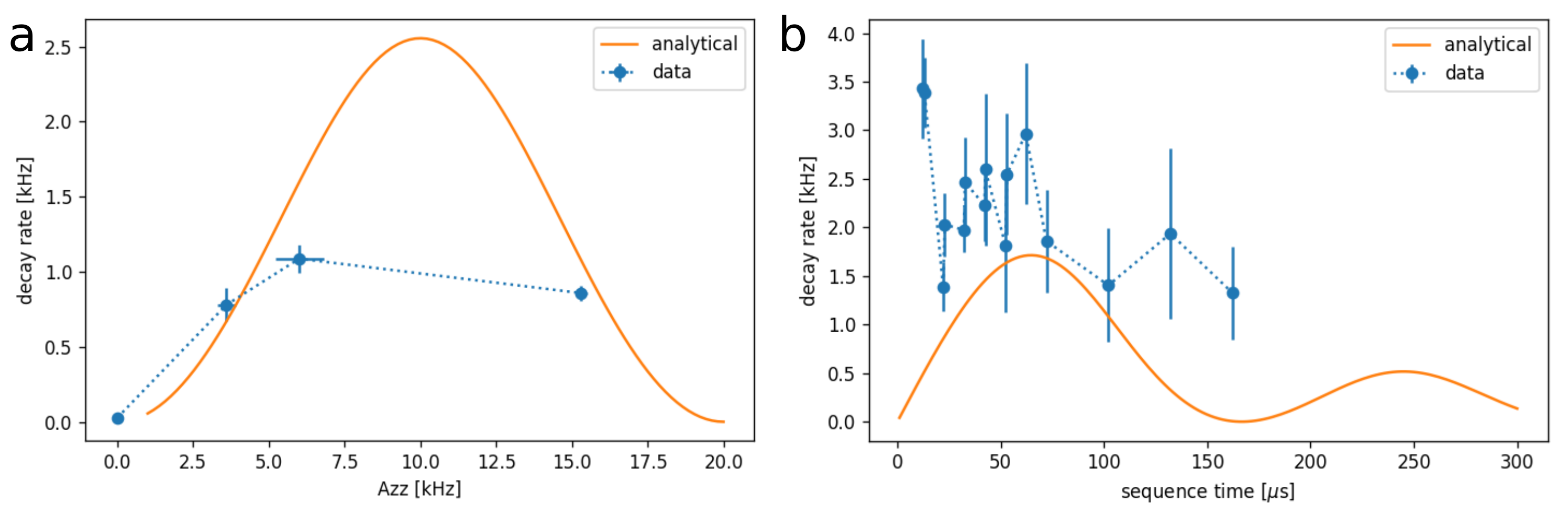}
\captionof{figure}{\label{fig:model-linewidth} \textbf{Contribution of decay rate by initialization infidelity of the sensor spin}. \textbf{a} Decay rates computed from equation \ref{eq:gamma-infidelity} for $\phi=2 \pi \cdot 50 \mu \mathrm{s} \cdot A_{zz}$ and $f=0.94$. Experimental data for the NV-$^{13}$C pairs from figure \ref{fig:Linewidth-discussion} in the main text. \textbf{b} Computed analytical decay rate for $A_{zz}=5 \, \mathrm{kHz}$ and same infidelity. The experimental data is overlayed from NV$\#$5 measured with conventional qdyne (kddxy). }
\end{figure*}

Consecutive readouts during the free precession of the nuclear target spin undoubtedly results in a charge state and spin state shuffling. In the following we discuss the decoherence based on a spin state initialization infidelity of the NV$^-$ and NV$^0$. First, we limit our discussion for sampling times in the range of $50 \, - \, 300 \, \mu \mathrm{s}$, relevant for the double qdyne. When we compare this time range to the  NV$^0$ spin lifetime $T_1^0 \approx 10 \, \mu \mathrm{s}$ we see that the lifetime is much shorter then the sampling time. This means we can neglect the NV$^0$ initialization infidelity as the spin state is changing multiple times during the time scales between two measurements. However, for the  NV$^-$ the spin lifetime $T_1^- = 1.2 \, \mathrm{ms}$ is much larger then the sampling time which leads to a initialization infidelity dominated dynamics. The NV$^-$ initial density matrix $\rho_{NV^-}$ when included the infidelity is given by:
\begin{equation}
	\rho_{NV^-} = (1 - f) \cdot \mathds{1} + (2f-1) \ket{0}\bra{0},
\end{equation}
where $f$ is the initialization fidelity and we assumed the nuclear spin is initially polarized in state $\ket{0}$. To study the dynamics of the system we assume a superposition state for the target nuclear spin of $\rho_{^{13}\mathrm{C}} = \frac{1}{2}(\mathds{1} + \sigma_x)$. The combined systems evolve under the Hamiltonian $H = A_{zz} S_z I_z = \frac{A_{zz}}{4}\sigma_z \otimes \sigma_z$ for the time $\tau$ between the measurements. The final density matrix of the target spin is computed from $\rho^\prime_{^{13}\mathrm{C}}  = \mathrm{Tr}_{NV^-} \left[U^\dagger (\rho_{NV^-} \otimes \rho_{^{13}\mathrm{C}}) U\right])$ and given by:
\begin{equation}
	\rho_{^{13}\mathrm{C}} = \frac{1}{2} \mathds{1} + \cos(\phi) \sigma_x + (2f-1) \sin(\phi) \sigma_y,
\end{equation}
where $\phi = A_{zz} \tau$ is the phase acquired for the time between two measurements. When we consider only the amplitude of the transverse component $\langle I_{\mathrm{trans.}} \rangle = \sqrt{\langle I_x\rangle^2 + \langle I_y\rangle^2}$, we get:
\begin{equation}
	\langle I_{\mathrm{trans.}} \rangle = \sqrt{\cos^2(\phi) + (2f-1)^2\sin^2(\phi)} \leq 1,
\end{equation}
which reduces the transverse component with each evolution cycle. The decay rate $\Gamma$ is derived by taking the logarithm and dividing by $\tau$ which leads to:
\begin{equation}
\label{eq:gamma-infidelity}
\begin{split}
	- \Gamma &= \frac{1}{2} \log\left( \cos^2 \phi + (2f-1)^2 \sin^2\phi \right) /\tau\\
	&= 2 f(1-f) \sin^2\phi / \tau + O((1-f)^2),
\end{split}
\end{equation}
where we assumed $(1-f)$ to be small for the Taylor expansion. \newline
Experimentally, we measured up to a time between two measurements of $10 - 200 \, \mu \mathrm{s}$. The spin initialization fidelity of NV$^-$ into $\ket{0}$ is commonly estimated to be 0.9 and the probability to be in the NV$^-$ charge state is 0.7. The analytical figures \ref{fig:model-linewidth} a $\&$ b were computed with a fidelity of $f=0.94$. From this computation we show that the decoherence based on the infidelity is plausible and in the right order of magnitude. However, when we tried to reproduce the destinct oscillatory feature by increasing the time between readouts, we could not reproduce them. A possible explanation can be that more than one mechanism is contributing to the decoherence, e.g. the decoherence based on NV$^-$ spin state relaxtion $T_1$, and is a topic of further investigations.

\section{Sensitivity Estimation}

\subsection{Nuclear precession frequency uncertainty scaling}
\begin{figure}[h!]
\centering
\includegraphics[width = 0.7 \textwidth]{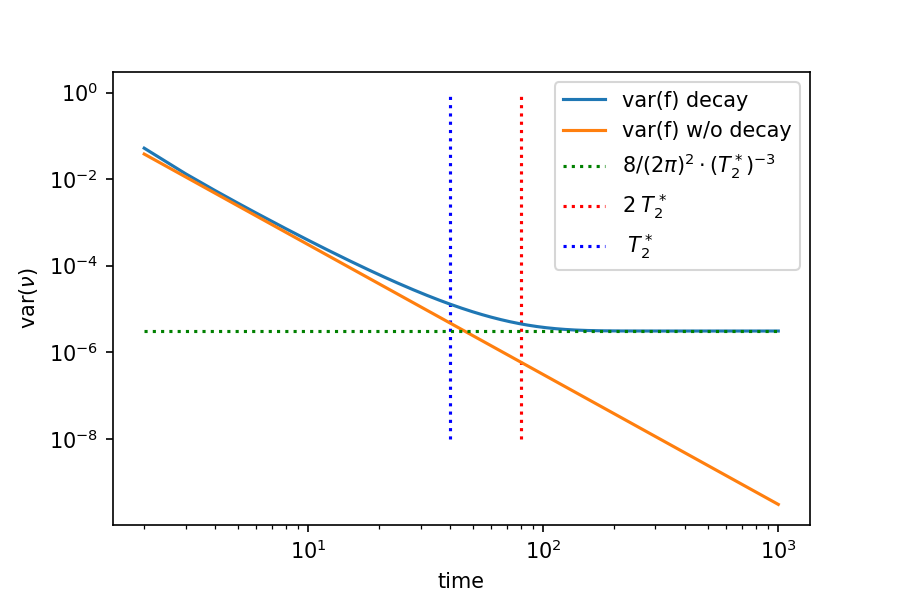}
\caption{\textbf{Frequency sensitivity estimation scaling for a decaying oscillating signal.} The qualitative signal's frequency variance scales with time $t$ to the power of $-3$ (orange). For short times a decaying signal (blue) follows the same power law, but the decay limits the relative variance to $8/(2\pi)^2 \cdot (T_2^*)^{-3}$. In this qualitative figure the decay time $T_2^*$ was set to 40 and both ($T_2^* \, \& \,2\, T_2^*$) are shown as vertical lines.}
\label{fig:varf}
\end{figure}

In the following we want to derive a measure of the sensitivity for NV based nuclear spin precession signals. In general a NMR signal $s$ is given by:
\begin{equation}
    s[n] = A \cdot \sin(2 \pi \nu n + \phi)\cdot  \exp(-\beta t) + w[n],
\end{equation}
where $n$ describes the $n$-th data point of the time domain signal recorded with constant sampling intervals $\Delta T$. The signal is described by the amplitude $A$, frequency $\nu$, initial phase $\phi$, decay constant $\beta = \Delta T/ T_2^*$ and $w(n)$ the white noise contribution at each data point. In NMR spectroscopy we focus on the frequency estimation of the signal, where we follow the examples 3.5 and 3.14 in reference \cite{sengijpta1995fundamentals} and reference supplemental material \cite{gemmel2010ultra}  for the exponential decay correction.
The variance of the estimated frequency can be determined by the Cramer-Rao lower bound from the Fisher information matrix. The final result for the variance of $\nu$ in reference \cite{gemmel2010ultra} leads to:
\begin{equation}
\label{eq:varf}
    \rm{var}(\nu) = \frac{12}{(2\pi)^2(A/\rho_\alpha)^2 T^3}C[T],
\end{equation}
where $\rho_\alpha$ is the power spectral density and $C[T]$ is a factor including the exponential decay. If $C[T]=1$ the result is equal to the non-decaying sine wave frequency estimation as in example 3.14\cite{sengijpta1995fundamentals}. However because of the decay the frequency resolution is bound by the decay and $C[T]$ is given by:
\begin{equation}
    C[T] = \frac{T^3}{12(\Delta T)^3} \frac{(1-z^2)^3 (1-z^{2N})}{z^2 ( 1- z^{2N})^2 - N^2 z^{2N}(1-z^2)^2},
\end{equation}
with $z = \exp(-\zeta) \approx 1 - \zeta$ and $N= T/\Delta T$ the number of data points. Taking the limit $N\rightarrow \infty$ for small $\zeta,\zeta^2 \approx 0$ we can simplify $C$ to:
\begin{equation}
    C =  \frac{T^3}{12(\Delta T)^3} (2\zeta)^3 = \frac{8T^3}{12 \left(T_2^*\right)^3}
\end{equation}
finally we get the overall variance bound of the decaying sine of:
\begin{equation}
\label{eq:fbound}
    \rm{var}(\nu) = \frac{8}{(2\pi)^2(A/\rho_\alpha)^2 (T_2^*)^3}.
\end{equation}
The variance of the signal frequency (equation \ref{eq:varf}) is plotted in figure \ref{fig:varf} for the parameters ($A/\rho_\alpha=1$, $T_2^*=40$, $\Delta T=1$, $T=[1,1000]$). 
We see that the variance quickly saturates after $2\; T_2^*$ and eventually reaches the variance bound (equation \ref{eq:fbound}). 
Assuming that that white noise originates from sensor itself, e.g. as spin projection noise, and photon counting statistics (shot-noise), it is natural to link sensor sensitivity with noise power spectral density $\eta = \sigma(B) \sqrt{T} \approx \rho_\alpha$.

\subsection{Sensor sensitivity for double resonance heterodyne sensing}
\label{supp:sensitivity}
Our NMR protocol relies on the sensor-target interaction and the free evolution of the target. The sensitivity of the sensor can be estimated from previous work for CSTE based protocols relying on $T_2$ coherence time and is typically $\eta^{T_2}_{\mathrm{NV}} \approx 10-30 \, \mathrm{pT}/\sqrt{\mathrm{Hz}}$ for ensemble diamonds \cite{glenn2018high} and $\eta^{T_2}_{nv} \approx 136 \, \mathrm{nT/\sqrt{Hz}}$ for shallow implanted 10-20 nm near surface NV centers \cite{WATANABE2021294}. Because we have two parts of our sensing protocol we need to adjust the sensitivity with the total time between measurements $\Delta T = T_{\mathrm{sensing}} + T_{\mathrm{fid}} + T_{\mathrm{rf}}$, where $T_{\mathrm{sensing}}$ is the time of the full sensing part, $T_{\mathrm{fid}}$ is the free precession time of the target system and $T_{\mathrm{rf}}$ is the time needed for the transition between the two. This leads to a reduced effective sensitivity of the sensor $\eta_{\mathrm{eff.}}$:
\begin{equation}
	\eta_{\mathrm{eff.}} = \eta_{\mathrm{NV}} \cdot \sqrt{\frac{\Delta T}{T_{\mathrm{sensing}}}}.
\end{equation}
We find that there is a trade-off ratio between the sensing time and the fid time when we consider an optimized frequency estimation uncertainty of the target system. 

\subsection{Balance between sensing and precession time}
\label{supp:optimal-ratio}
Finally we calculate the optimal ratio between sensing and free precession time. As described above the sensitivity of the sensor scales with $\sqrt{ \Delta T  / T_{\mathrm{sensing}}}$ because of the time overhead from the free evolution and change of basis. 
But also the sensor target interaction in our protocol stops the precession and therefore reduces the effective spin precession time. This reduces the relative frequency uncertainty $\sigma_\omega/\omega$ (typically for NMR measured in part per million). 
When the precession is interrupted the frequency reduces effectively by $T_{\mathrm{fid}} / \Delta T$.\newline
In summary the relative frequency  uncertainty is given by:
\begin{equation}
	\frac{\sigma_\omega}{\omega_{\mathrm{eff.}}} =  \frac{2\sqrt{2}}{\gamma_{\rm{eff.}}\, B} \frac{\eta_{\rm{eff}}}{\rm{A[B]} } \frac{1}{(T_2^*)^{3/2}}
\end{equation}
where only $\gamma_{\rm{eff.}}$ and $\eta_{\rm{eff}}$ depend on $\Delta T$. We take the derivative of the relevant part of the frequency uncertainty:
\begin{equation}
\begin{split}
	\frac{\rm{d}}{\rm{d}\Delta T} \frac{\sqrt{\Delta T}}{(1 - T_{\rm{meas}}/\Delta T)} &=   \frac{\left(\Delta T - 3 T_{\rm{meas}}\right)\sqrt{\Delta T}}{2(\Delta T - T_{\rm{meas}})^2},
\end{split}
\end{equation} 
with $T_{\rm{meas}} = T_{\mathrm{sensing}} +  T_{\mathrm{rf}}$. The above equation has a zero crossing at $\Delta T = 3 T_{\rm{meas}}$, therefore the ideal ratio is 2:1 of free precession time to measurement time. 
\subsection{Sensitivity of double resonance heterodyne detection under ideal condition}
 In the following we calculate the final sensitivity for our single NV proof of principle experiment and also for an ideal experiment with an ensemble of NV centers. In table \ref{tab:nv-parameters} we show the parameters for our experiment and for an ideal NV ensemble experiment and the resulting effective sensitivity $\eta_{\mathrm{eff.}}$.
 
\begin{table}[h!]
    \centering
    \begin{tabular}{|c|c|c|c|}
    \hline
    Parameter & Single NV deep  &     Single NV shallow  & NV ensemble \\
    &  (present work) & (d=10 nm, ideal) & (ideal) \\
    \hline
    $B_0, T$ & $0.25$ & 3 \cite{aslam2017nanoscale} & 3  \cite{aslam2017nanoscale}\\
   $T^2_{\mathrm{sens}}$ & $\left(T_2^* \, \mathrm{limit}\right)$ 50 $\mu$s & $\left(T_2 \, \mathrm{limit}\right)$ 300 $\mu$s   \cite{WATANABE2021294} & $\left(T_2 \, \mathrm{limit}\right)$ 10 $\mu$s \cite{glenn2018high,Masuyama_2018}\\

   $\Omega^{^{13}C}_{\mathrm{rabi}}$ & 15 kHz & 250 kHz \cite{herb2020high} &250 kHz \cite{herb2020high} \\

   $\eta_{\mathrm{NV}}$ & $\approx 900 \, \mathrm{nT/\sqrt{Hz}}$ &$\approx 140 \, \mathrm{nT/\sqrt{Hz}}$ &$30 \, \mathrm{pT/\sqrt{Hz}}$ \cite{glenn2018high}\\

   $T_{\mathrm{sensing}}$ & $60 \, \mathrm{\mu s}$ &  $300 \, \mathrm{\mu s}$ & 10 $\mathrm{\mu s}$\\
   $\Delta T_{\mathrm{opt.}}$ & 390 $\mu$s & 912 $\mathrm{\mu s}$& 42 $\mu$s\\
   $\sqrt{\Delta T_{\mathrm{opt.}} / T_{\mathrm{sensing}}}$ & 2.549 & 1.74 &2.0 \\
   \hline
   $\eta_{\mathrm{eff.}}$ &  $2.3 \, \mathrm{\mu T/\sqrt{Hz}}$& $243 \, \mathrm{nT/\sqrt{Hz}}$ & $60 \, \mathrm{pT/\sqrt{Hz}}$\\
   \hline
    \end{tabular}
    \caption{Parameters for our experiment and an ideal realization of the proposed experiment}
    \label{tab:nv-parameters}
\end{table}

\subsection{Summary of the sensitivity estimation}
The above discussion showed how the variance of the frequency estimation scales with  the measurement time and  the sensor sensitivity. The earlier shows the motivation for heterodyne sensing as the $\mathrm{var}(\nu)$ scales with $T^{-3}$ for short times and is ultimately saturated at $T=2T_{2,n}^*$. This lead to improved sensitivity of heterodyne protocols compared to absorption based experiments (e.g. also discussed in reference \cite{schmitt2017submillihertz} figure 4 d). The later determines the sensitivity of our quantum sensor when applied in the NMR protocol. Because of the rf manipulation and the free precession time overhead the sensitivity is reduced compared to magnetic field sensing and eventually given by 2.3 $\mathrm{\mu T/\sqrt{Hz}}$ for our single NV and 243 (60) $\mathrm{nT (pT)/\sqrt{Hz}}$ for an ideal (ensemble) experiment.

\begin{figure}[htbp]
\begin{center}
\includegraphics[width = 0.45\columnwidth]{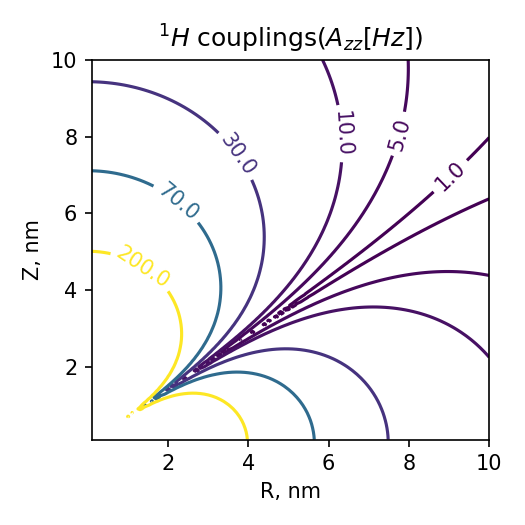}
\includegraphics[width = 0.45\columnwidth]{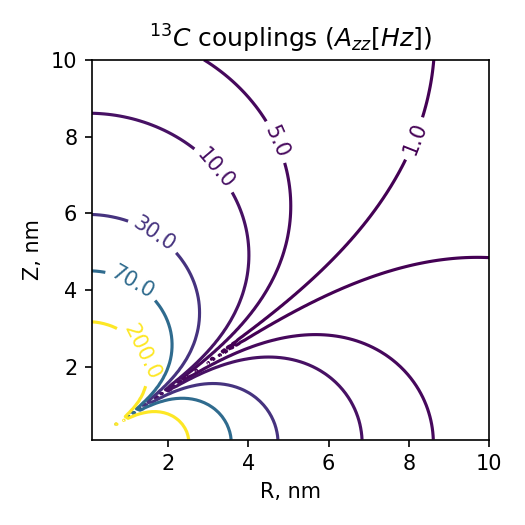}
\caption{Estimation of the dipolar-dipolar coupling for the hyperfine splitting of $A_{zz}$ for the $^1H$ (left) and $^{13}C$ nuclei (right)}
\label{fig:distance}
\end{center}
\end{figure}

\newpage

\newpage 
\section{Variables}
Variables used in the main text and in the supplemental material.
\begin{table}[h!]
    \centering
    \begin{tabular}{|c|c|}
    \hline
    Variable     & Description\\
    \hline
    $\omega_L$     & Larmor frequency nuclear spin \\
    $B$			   & Magnetic field \\
    $\gamma_N$		& Nuclear gyromagnetic ratio \\
    $I_{x,y,z}$     & Nuclear spin operator \\
    $S_{x,y,z}$     & Sensor/electron spin operator \\
    \hline
    $ A_{zx} $& Hyperfine interaction between $S_z$ and $I_x$ [kHz] \\
    $ A_{zz} $& Hyperfine interaction between $S_z$ and $I_z$ [kHz] \\
    $\tau$ & FID time in the theoretical description \\ 
    $\tau_{zz}$ & Sensing interrogation time for $A_{zz}$ interaction\\
    $\tau_1$ & free precession time in fig. 2a experiment\\
    \hline
    $\tau_2$ & simulated interaction time in fig. 2a experiment\\
    $\gamma_{eff}$ & Effective reduced g-factor of the nuclear spins \\
    $\alpha$ & Measurement strength\\
    $N$ & number of single shot readout repetitions \\
    $M$ & Number of repetitions of the measurement sequence \\
    \hline
    $T_2^*$ & Free precession decay time\\
    $T_1$ & Electron longitudinal relaxation time (NV$^-$ and NV$^0$ indicated)\\
    $T_{seq} $ & sequence duration\\
        $\nu$ & \\
        $\Gamma$ & decay rate of the nuclear spin signal \\
    \hline
    $\Phi$ &  \\
    $\Omega$ & Rabi frequency ($i:$ nuclear, $^{14}$N, $^{13}$C, $s:$ sensor)  \\
    $\Lambda$ & Effective Rabi frequency when included off-resonance\\
    $\Delta$ & frequency difference between reference and spin transition frequency [kHz]\\
    $\epsilon$ & Amplitude pulse error ($\%$) \\
    \hline
    $T_{rf}, T_{nuc}$ & Time for nuclear spin control pulses\\
    $A, \rho_\alpha$ & Signal in T, noise power spectral density, $\mathrm{T/\sqrt{Hz}}$ \\
        $\sigma(\nu)$ & standard deviation for the frequency estimation\\
        $T_{\mathrm{meas}}, T_{\mathrm{rf}}, T_{\mathrm{fid}}, T_{sens}$ & \\
        $ f$& fidelity of electron spin initialization by the green pulse \\
        \hline
        $ \eta$& sensitivity of the electron spin to magnetic field [$\mathrm{T}/\sqrt{\mathrm{Hz}}$]\\
        $\beta$ & Precession angle between pi/2 rf-pulses \\
        $\Delta T_{(opt)}, T_{total}, $ & (Optimum) sampling intervals\\
        $\tau_{int}(App. B3)$ & Sensing interaction time general (kddxy or ENDOR)\\
    \hline
    \end{tabular}
    \caption{Table of used variables}
    \label{tab:my_label}
\end{table}

\newpage

\end{document}